\documentclass[reprint,superscriptaddress,showpacs,amsmath,amssymb,floatfix,aps,pdftex]{revtex4-1}
\usepackage[hiresbb]{graphicx}% Include figure files
\usepackage{dcolumn}% Align table columns on decimal point
\usepackage{bm}% bold math

\usepackage{color}
\usepackage[super]{nth}

\begin{document}
%%%%% Title %%%%%
\title{
Spin correlations of quantum-spin-liquid and quadrupole-ordered states of Tb$_{2+x}$Ti$_{2-x}$O$_{7+y}$
}
%%%%%%%%%%%%%%%%%

%%%%% Authors %%%%%
\author{Hiroaki Kadowaki}
\affiliation{Department of Physics, Tokyo Metropolitan University, Hachioji, Tokyo 192-0397, Japan}

\author{Mika Wakita}
\affiliation{Department of Physics, Tokyo Metropolitan University, Hachioji, Tokyo 192-0397, Japan}

\author{Bj\"{o}rn F{\aa}k}
\affiliation{Institut Laue-Langevin, CS 20156, 38042 Grenoble Cedex 9, France}

\author{Jacques Ollivier}
\affiliation{Institut Laue-Langevin, CS 20156, 38042 Grenoble Cedex 9, France}

\author{Seiko Ohira-Kawamura}
\affiliation{Neutron Science Section, MLF, J-PARC Center, Shirakata, Tokai, Ibaraki 319-1195, Japan}

\author{Kenji Nakajima}
\affiliation{Neutron Science Section, MLF, J-PARC Center, Shirakata, Tokai, Ibaraki 319-1195, Japan}

\author{Jeffrey W. Lynn}
\affiliation{NIST Center for Neutron Research, National Institute of Standards and Technology, Gaithersburg, Maryland 20899-6102, USA}

%%%%%%%%%%%%%%%%%%%
\date{\today}

%%%%% Abstract %%%%%
\begin{abstract}
Spin correlations of the frustrated pyrochlore oxide Tb$_{2+x}$Ti$_{2-x}$O$_{7+y}$  
have been investigated by using inelastic neutron scattering 
on single crystalline samples ($x=-0.007, 0.000,$ and $0.003$), which have 
the putative quantum-spin-liquid (QSL) 
or electric-quadrupolar ground states. 
Spin correlations, which are notably observed in nominally elastic scattering, 
show short-ranged correlations around $L$ points [$\bm{q} = (\tfrac{1}{2},\tfrac{1}{2},\tfrac{1}{2})$], 
tiny antiferromagnetic Bragg scattering at $L$ and $\Gamma$ points, 
and pinch-point type structures around $\Gamma$ points. 
The short-ranged spin correlations were analyzed 
using a random phase approximation (RPA) 
assuming the paramagnetic state and two-spin interactions among Ising spins. 
These analyses have shown that the RPA scattering intensity 
well reproduces the experimental data 
using temperature and 
$x$ dependent coupling constants of up to \nth{10} 
neighbor site pairs. 
This suggests that no symmetry breaking occurs in the QSL sample, 
and that a quantum treatment beyond the semi-classical RPA approach is required. 
Implications of the experimental data and 
the RPA analyses are discussed. 
\end{abstract}
%%%%%%%%%%%%%%%%%%%%

%\pacs{}
\maketitle

\section{Introduction}
Geometrically frustrated magnets archetypally on 
the two-dimensional (2D) triangle \cite{Wannier50} and kagom\'{e} \cite{Shyozi51,Qi2008} lattices, 
and on the three-dimensional (3D) pyrochlore lattice \cite{Gardner10} 
have been actively studied for decades \cite{Lacroix11}. 
Among classical frustrated magnets, 
spin ice \cite{Bramwell01} has been extensively studied 
from many viewpoints, e.g., 
macroscopically degenerate ground states \cite{Ramirez99}, 
partial lifting of the degeneracy under magnetic field \cite{MatsuhiraJPCM2002}, 
and fractionalized excitations \cite{Castelnovo08,Kadowaki09}. 
Quantum effects in frustrated magnetic systems 
ranging from quantum annealing \cite{KadowakiNishimori1998,King2018} 
to quantum spin liquid (QSL) states \cite{Savary2017}, 
the origin of which dates back to the proposal of the RVB state \cite{Anderson73}, 
have attracted much attention. 
Experimental challenges of finding 
real QSL substances \cite{Hirakawa1985,Gardner99} 
and of investigating QSL states using available techniques 
\cite{Han2012,Ross11,Chang12,Shen2016,Fak2017,Sibille2018} 
have been addressed in recent years. 

Among frustrated magnetic pyrochlore oxides \cite{Gardner10} 
a non-Kramers pyrochlore magnet 
Tb$_{2+x}$Ti$_{2-x}$O$_{7+y}$ (TTO) \cite{Kadowaki2018} 
has been investigated for decades as a QSL candidate, 
since conventional magnetic order has not been observed 
in any experiments under zero field and zero static pressure \cite{Gardner99,Gardner10}. 
On the basis of 
theoretical insight that TTO is not much different from classical spin ice, 
the phrase quantum spin ice (QSI) was coined for the QSL state of TTO \cite{Molavian07,Gingras14}. 
However, its nature has remained elusive. 
Recently we showed that this putative QSL state 
is limited in a range of the small off-stoichiometry parameter 
$x < x_{\text{c}} \simeq -0.0025$ \cite{Taniguchi13,Wakita2016,Kadowaki2018}. 
In the other range $x_{\text{c}} < x$, we showed that 
TTO undergoes a phase transition most likely to 
an electric multipolar [or quadrupole ordered (QO)] state ($T<T_{\text{c}}$) \cite{Takatsu2016prl,Takatsu2016JPCS,Kadowaki2018prb}, 
which is described by a pseudospin-$\frac{1}{2}$ Hamiltonian 
modified from the classical spin ice to a quantum model 
by adding transverse pseudospin terms \cite{Onoda11}. 
The estimated parameter set of this Hamiltonian \cite{Takatsu2016prl} 
is close to the theoretical phase boundary 
between the electric quadrupolar state and a U(1) QSL state (QSI) \cite{Lee12,Hermele04}, 
which is thereby a theoretical QSL candidate for TTO. 
At present, few researchers have addressed the problem of the QSL state of TTO 
using well $x$-controlled samples. 

Previous neutron scattering experiments on 
TTO, which were performed 
on samples with unknown and known $x$, 
showed that spin correlations, defined by the 
wavevector dependence of 
scattering intensity are most clearly seen in 
energy-resolution-limited (nominally) elastic scattering 
at low temperatures. 
In the observed spin correlations there are three important features: 
magnetic short-range order (SRO) with the wavevector 
$\bm{q} = (\tfrac{1}{2},\tfrac{1}{2},\tfrac{1}{2})$ 
($L$ point of the first Brillouin zone of the FCC lattice)
\cite{Yasui2002,Fennell2012,Petit12,Fritsch13}, 
pinch point structures around $\bm{q} = 0$ ($\Gamma$ point) \cite{Fennell2012,Petit12}, 
and tiny antiferromagnetic Bragg reflections 
at $L$ and $\Gamma$ points \cite{Taniguchi13,Takatsu2016prl}. 
It should be noted that details of the observed scattering intensities 
in these studies depended on samples (on $x$). 
This may intriguingly suggest that 
the ground states of TTO are potentially highly degenerate 
and they are lifted in various ways depending on slight 
differences of samples. 

Very recently we performed inelastic neutron scattering (INS) experiments 
on $x$-controlled TTO single-crystalline samples 
with $x=-0.007< x_{\text{c}}$ (QSL) and $x_{\text{c}} < x=0.000, 0.003 $ (QO) \cite{Kadowaki2018}. 
In this paper we focus on the 
$\bm{q} = (\tfrac{1}{2},\tfrac{1}{2},\tfrac{1}{2})$ SRO of these samples 
and perform quantitative analyses 
in order to shed light on how these spin correlations reflect the QSL state. 
In previous investigations \cite{Fritsch13,Guitteny2015}, 
analyses of the $\bm{q} = (\tfrac{1}{2},\tfrac{1}{2},\tfrac{1}{2})$ SRO 
were carried out by assuming that there exist 
static short-ranged classical spins with cluster sizes of the order 10 {\AA}. 
However no clusters which 
adequately reproduce the observed intensity pattern were found, 
although a few clusters showing limited goodness-of-fit were 
obtained \cite{Fritsch13,Guitteny2015}. 
This failure indicates either that the samples were not well controlled 
or that the analysis methods they used are not sufficiently systematic. 

The first problem of controlling the composition of the samples 
is resolved in the present study. 
In contrast, the second problem can originate from a profound property of the QSL state, 
and will be resolved only by analyses 
reflecting the quantum nature of the many-body ground state. 
However, since no practical quantum model calculations are available at present, 
in the present study, 
we attempt to apply a systematic but still semi-classical approach 
using a random phase approximation (RPA) \cite{Jensen91}. 
This would lead us to a reasonable result if the SRO could be 
interpreted within the classical spin paradigm, 
or leads us to a certain paradoxical result 
if it essentially contains many-body quantum effects.

\section{Methods}
\subsection{Experimental Methods}
Single crystalline samples of Tb$_{2+x}$Ti$_{2-x}$O$_{7+y}$ 
with $x=-0.007, 0.000$ and $0.003$ 
used in this study are those of Ref.~\cite{Kadowaki2018}, 
where methods of the sample preparation and the estimation of $x$ 
are described. 
The QSL sample with $x=-0.007$ remains in the paramagnetic 
state down to 0.1 K.
The QO samples with $x=0.000$ and $x=0.003$ very likely have small and 
large electric quadrupole orders, respectively, 
at $T \ll T_c \sim 0.4 $ K \cite{Taniguchi13,Wakita2016}. 
We note that the values of $x$ among different investigation 
groups are not necessarily consistent \cite{Kadowaki2018}, 
and that our $x$ values of the samples used in 
Refs.~\cite{Taniguchi13,Kadowaki2015,Takatsu2016prl,Takatsu2016JPCS,Wakita2016,Takatsu2017,Kadowaki2018}
are self-consistent. 

Neutron scattering experiments were carried out on the time-of-flight (TOF) spectrometer 
IN5 \cite{Fak2015,Fak2016} 
operated with $\lambda = 8$ {\AA} at ILL for the $x = -0.007$ and 0.000 crystal samples. 
The energy resolution of this condition was $\Delta E = 0.021$ meV (FWHM) 
at the elastic position. 
Neutron scattering experiments for the $x = 0.003$ crystal sample were performed on the TOF 
spectrometer AMATERAS operated 
with $\lambda = 7$ {\AA} at J-PARC. 
The energy resolution of this condition was $\Delta E = 0.024$ meV (FWHM) at the elastic position. 
Each crystal sample was mounted in a dilution refrigerator 
so as to coincide its $(h,h,l)$ plane with 
the horizontal scattering plane of the spectrometer. 
The observed intensity data were 
corrected for background and absorption using a home-made program \cite{Kadowaki2018github}. 
Construction of four dimensional $S(\bm{Q},E)$ data object 
from a set of the TOF data taken by rotating each crystal sample 
was performed using {H}{\footnotesize ORACE} \cite{Horace2016}. 

To analyze the $\bm{Q}$-dependence of 
the (nominally) elastic scattering intensity (Fig.~1 in Ref.~\cite{Kadowaki2018}), 
we integrated $S(\bm{Q},E)$ in a small energy range $-\epsilon < E < \epsilon$. 
We chose $\epsilon = 0.025$ and $0.030$ meV for IN5 and AMATERAS data, respectively, 
which are a little larger than the instrumental resolutions. 
These 3D data sets $[S(\bm{Q})]_{\text{el}}= \int_{-\epsilon}^{\epsilon}S(\bm{Q},E)dE$ are normalized 
by the method described in Ref.~\cite{Kadowaki2018}, i.e., 
using the ``arb. units'' of Fig.~1 in Ref.~\cite{Kadowaki2018}. 
Consequently the elastic intensities 
can be compared mutually among the three samples. 

\subsection{RPA model calculation}
The RPA model calculation of $S(\bm{Q},E)$ 
using the pseudospin-$\frac{1}{2}$ Hamiltonian 
appropriate for quadrupole ordered phases 
is described in Ref.~\cite{Kadowaki2015}. 
We used a similar RPA method to calculate 
the elastic scattering intensity $[S(\bm{Q})]_{\text{el}}$ 
assuming that the system is in the paramagnetic phase. 
This assumption is made because we are interested mainly 
in the low-temperature QSL and the high-temperature paramagnetic states. 
Details and related definitions are described in 
Appendix~\ref{appendix_RPA}.

For the sake of simplicity we consider a pseudospin-$\frac{1}{2}$ Hamiltonian 
which is decoupled between magnetic dipole ($\sigma_{\bm{r}}^{z}$) 
and electric quadrupole ($\sigma_{\bm{r}}^{x}$ and $\sigma_{\bm{r}}^{y}$) terms, 
the latter of which can be neglected for the present purpose. 
We adopt a magnetic Hamiltonian expressed by 
\begin{eqnarray}
&& H_{\text{m}} = \sum_m J_m \left\{ \sum_{\langle {\bm r} , {\bm r}^{\prime} \rangle_m} 
\sigma_{\bm{r}}^{z} \sigma_{\bm{r}^{\prime}}^{z} \right\} + D r_{\text{nn}}^3 \nonumber \\
&\times& \sum_{\langle \bm{r} , \bm{r}^{\prime} \rangle} \left\{
\frac{ \bm{z}_{{\bm r}} \cdot \bm{z}_{{\bm r}^{\prime}} }{ | \Delta \bm{r} |^3} - 
 \frac{ 3 [\bm{z}_{{\bm r}} \cdot \Delta \bm{r} ] [\bm{z}_{{\bm r}^{\prime}} \cdot \Delta \bm{r} ] }{| \Delta \bm{r} |^5} \right\} \sigma_{\bm{r}}^{z} \sigma_{\bm{r}^{\prime}}^{z} \;,
\label{Hmag}
\end{eqnarray}
which is an expansion of that of Refs.~\cite{Kadowaki2015,Takatsu2016prl}. 
The first term of Eq.~(\ref{Hmag}) stands for 
magnetic coupling allowed by the space group symmetry between the Ising spin operators. 
The summation runs over coupling constants $J_m$ 
($m=1, \cdots , m_{\text{max}}$, $m_{\text{max}} \le 16$)
and corresponding site pairs $\langle {\bm r},{\bm r}^{\prime} \rangle_m$. 
These site pairs are listed in Table~\ref{table_J1J16}. 
The nearest-neighbor (NN) coupling constant $J_1$ is usually 
expressed as $J_{\text{nn}}$ for the NN spin ice model ($J_{\text{nn}}=J_1 >0$). 
The other couplings as far as 
\nth{10} neighbor site pairs 
had to be included to obtain good fit 
of the experimental data. 
Since the coupling constants beyond third-neighbor site pairs ($J_{m>4}$) 
are probably much smaller than $J_1$, 
they would be effective values or experimental parameters. 
The second term of Eq.~(\ref{Hmag}) represents the classical dipolar interaction \cite{Hertog00}, 
where $r_{\text{nn}}$ is the NN distance and $\Delta \bm{r} =  \bm{r} - \bm{r}^{\prime}$. 
The parameter $D$ is determined by 
the magnitude of the magnetic moment of the crystal field ground state doublet. 
We adopt $D = 0.29$ K, corresponding to the magnetic 
moment 4.6 $\mu_{\text{B}}$ \cite{Takatsu2016prl}.

The generalized susceptibility $\chi_{\nu,\nu^{\prime}}(\bm{k}, E=0)$ is computed by solving 
Eq.~(\ref{Gsus}) with $E=0$, i.e., 
\begin{equation}
\sum_{\nu^{\prime \prime}}[\delta_{\nu,\nu^{\prime \prime}} - \chi_{\text{L}} J_{\nu,\nu^{\prime \prime}}(\bm{k})] \chi_{\nu^{\prime \prime},\nu^{\prime}}(\bm{k}, 0) = \delta_{\nu,\nu^{\prime}} \chi_{\text{L}} \;, 
\label{GsusE0}
\end{equation}
where $J_{\nu,\nu^{\prime}}( \bm{k} )$ denotes 
the Fourier transform of the magnetic coupling constants [Eq.~(\ref{Ftransform})]
and $\chi_{\text{L}}$ is the local susceptibility [Eq.~(\ref{sus0})]. 
Using $\chi_{\nu,\nu^{\prime}}(\bm{k}, 0)$, 
the elastic scattering $[S(\bm{Q})]_{\text{el}}$ is given by 
\begin{eqnarray}
[S(\bm{Q}&=&\bm{G}+\bm{k})]_{\text{el}} \propto f(Q)^2
\sum_{\rho,\sigma,\nu,\nu^{\prime}} (\delta_{\rho,\sigma} - \hat{Q}_{\rho} \hat{Q}_{\sigma} ) \nonumber \\
& & \times U_{\rho,z}^{(\nu)} U_{\sigma,z}^{(\nu^{\prime})} 
\chi_{\nu,\nu^{\prime}}(\bm{k}, 0) \cos [ \bm{G} \cdot (\bm{d}_{\nu} - \bm{d}_{\nu^{\prime}}) ] \: ,
\label{integratedSQEelastic}
\end{eqnarray}
where $f(Q)$ is the form factor of Tb$^{3+}$,
in the quasi-elastic approximation [Eq.~(\ref{SQ_QEA})]. 

\section{Results}
\subsection{\label{results_IN5_QSL} QSL sample with $x=-0.007$}
\begin{figure*}[hbt]
\centering
\includegraphics[width=18.0cm,clip]{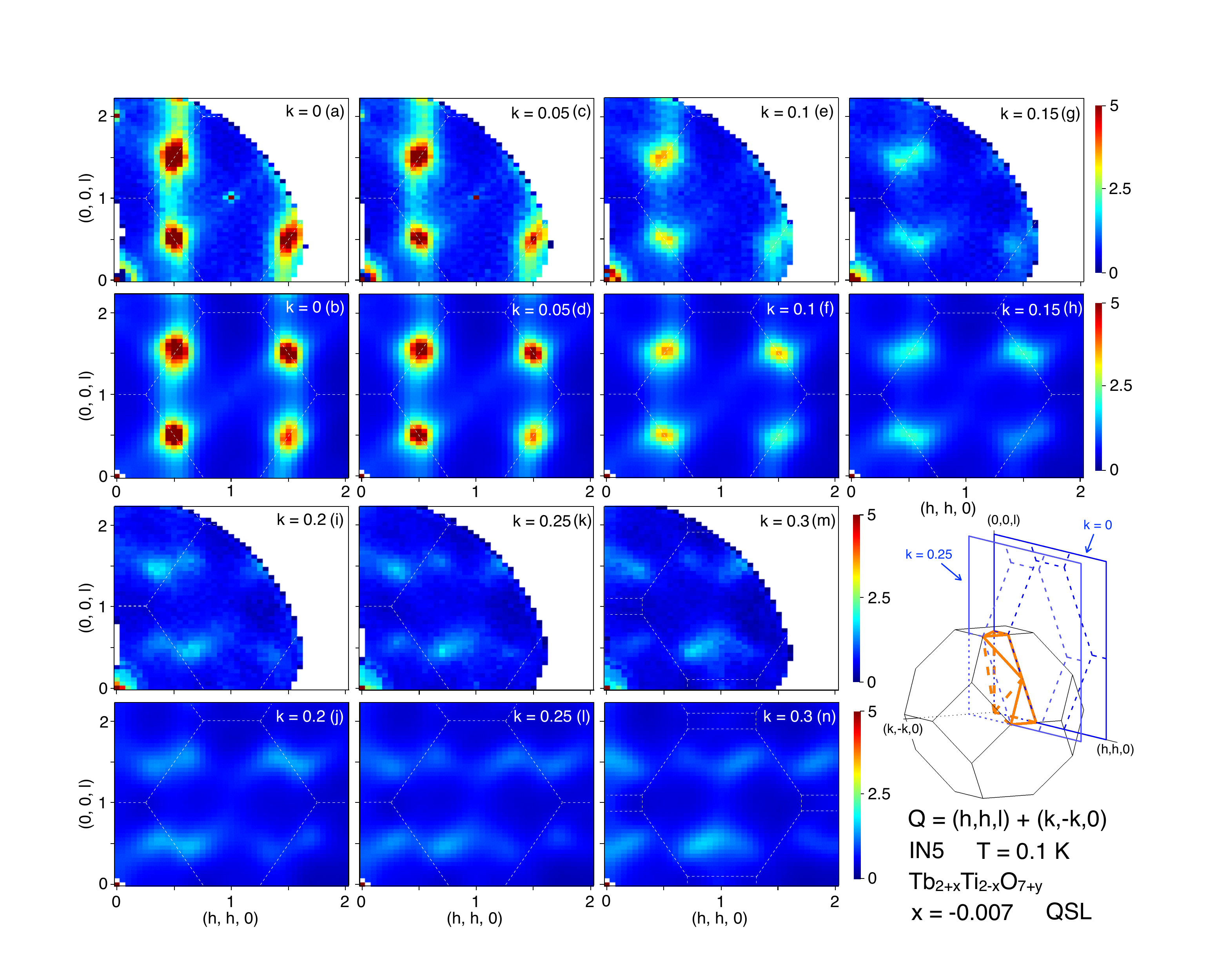}
\caption{ 
Intensity maps of 3D data $[S(\bm{Q})]_{\text{el}}$ 
taken at 0.1 K for the QSL sample with $x = -0.007$. 
The 3D data are viewed by 2D slices (a,c,e,g,i,k,m), which are parallel cross-sections 
of $\bm{Q}=(h,h,l)+(k,-k,0)$ with fixed $k$. 
These can be compared to the typical RPA $[S(\bm{Q})]_{\text{el}}$ (b,d,f,h,j,l,n) 
obtained by least squares fit 
using the 13 coupling constants, $J_1, \cdots, J_{13}$, 
listed in Table~\ref{FitJ1J14}. 
Dashed lines in these 2D slices (a-n) are boundaries of Brillouin zones. 
The bottom right corner shows the first Brillouin zone of 
the FCC lattice (thin black lines), 
irreducible zone (thick orange lines), 
and two 2D slice planes labeled 
$k=0$ and $0.25$ (blue lines). 
}
\label{QSL_IN5_Intensity_map0p1K}
\end{figure*}
\begin{figure}[htb]
\centering
\includegraphics[width=8.0cm,clip]{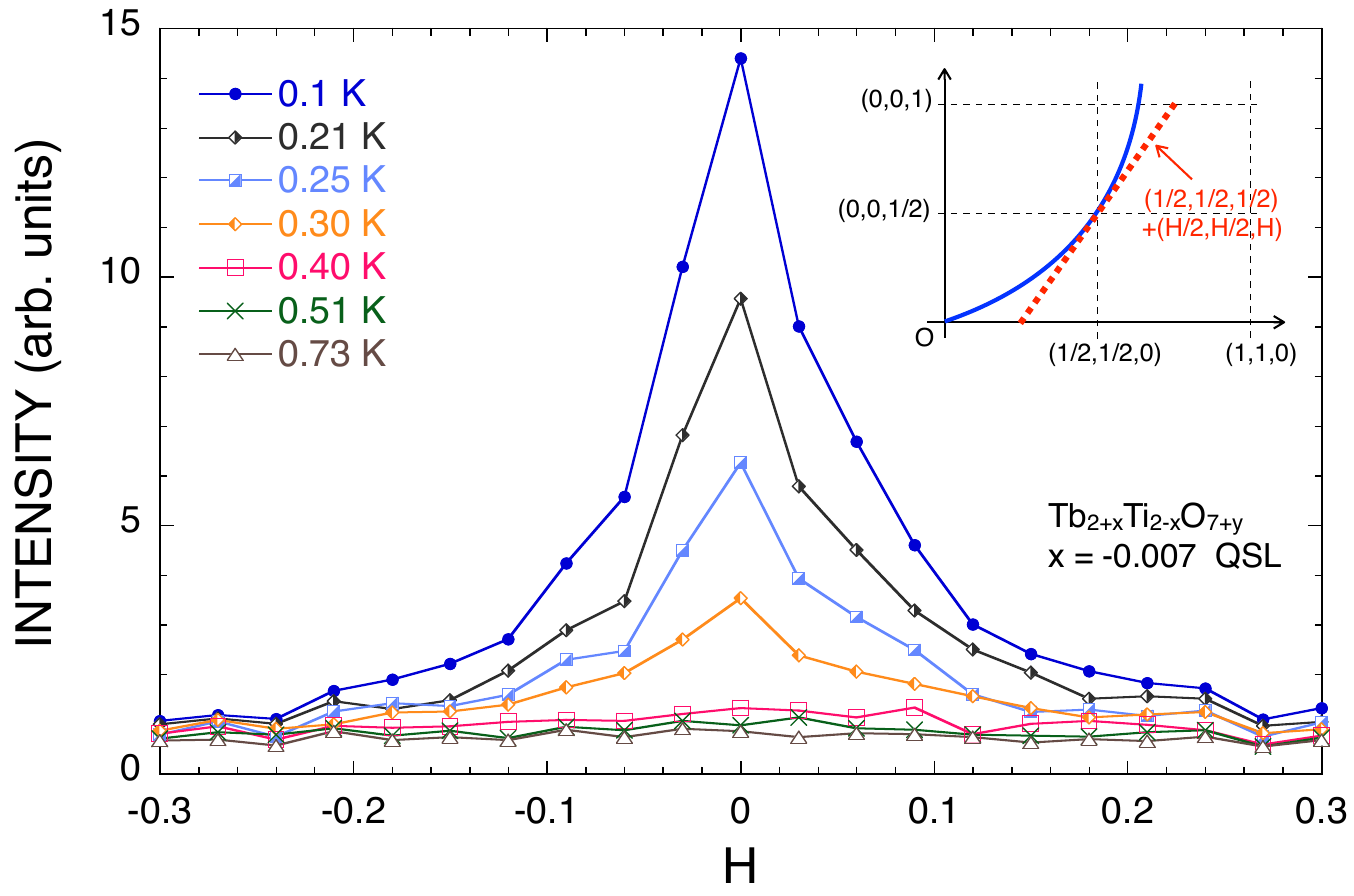}
\caption{ 
Temperature dependence of intensity $[S(\bm{Q})]_{\text{el}}$ 
along a trajectory through $\bm{Q}=(\tfrac{1}{2},\tfrac{1}{2},\tfrac{1}{2})$, 
which was measured by fixing the sample rotation angle. 
The abscissa of this figure is a projection 
of the $\bm{Q}$ trajectory (blue line in inset) 
to a straight line $\bm{Q}=(1/2,1/2,1/2)+(H/2,H/2,H)$ (red dashed line in inset). 
}
\label{QSL_HalfHalfHalf_Tdep}
\end{figure}
\begin{figure}[htb]
\centering
\includegraphics[width=8.0cm,clip]{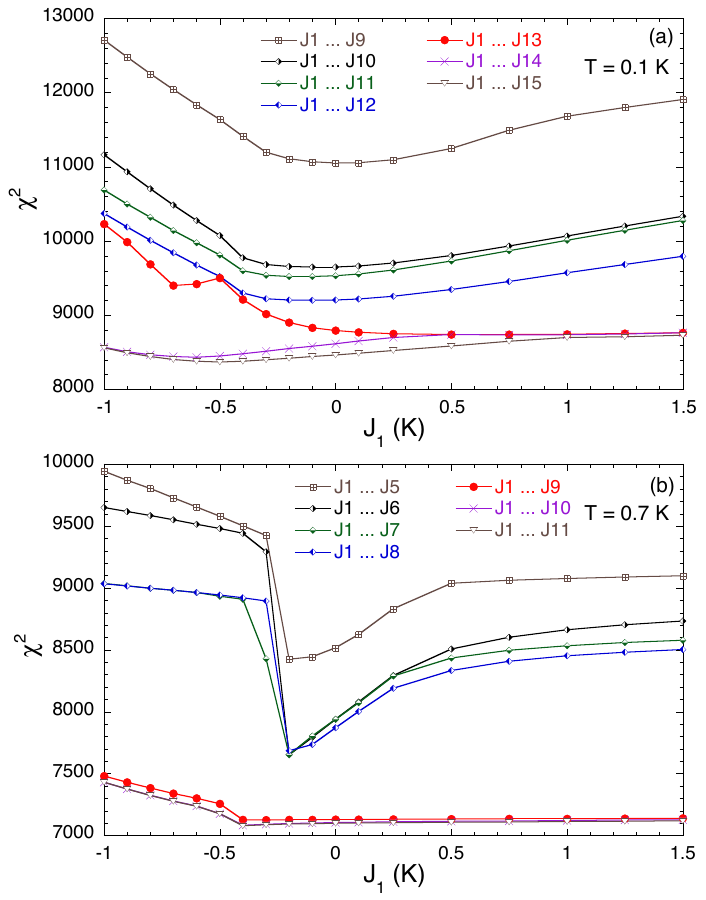}
\caption{ 
Minimized values of the weighted sum of squared residuals 
$\chi^2$ as a function of the fixed parameter $J_1$. 
(a) Results of 
least squares fits of $[S(\bm{Q})]_{\text{el}}$ with adjustable parameters 
$J_m$ ($m \le m_{\text{max}} = 9, \cdots , 15$)
for the QSL sample with $x=-0.007$ taken at $0.1$ K 
(Fig.~\ref{QSL_IN5_Intensity_map0p1K}). 
The number of fit data is 10185. 
(b) Results of  
least squares fits of $[S(\bm{Q})]_{\text{el}}$ with adjustable parameters 
$J_m$ ($m \le m_{\text{max}} = 5, \cdots , 11$)
for the QSL sample with $x=-0.007$ taken at $0.7$ K 
(Fig.~\ref{QSL_IN5_Intensity_map0p7K}).
The number of fit data is 10147. 
}
\label{chisq_QSL_IN5_0p1K0p7K}
\end{figure}
\begin{table*}
\caption{\label{FitJ1J14}
Typical coupling constants $J_m$ (in units of K) 
of Eq.~(\ref{Hmag}) obtained by least squares fits 
of observed 3D data sets $[S(\bm{Q})]_{\text{el}}$ to Eq.~(\ref{integratedSQEelastic}). 
The calculated $[S(\bm{Q})]_{\text{el}}$ using these $J_m$ are shown in 
Fig.~\ref{QSL_IN5_Intensity_map0p1K} ($x=-0.007$, $T=0.1$ K), 
Fig.~\ref{QSL_IN5_Intensity_map0p7K} ($x=-0.007$, $T=0.7$ K), 
Fig.~\ref{QO_IN5_Intensity_map0p1K} ($x=0.000$, $T=0.1$ K), 
Fig.~\ref{QO_IN5_Intensity_map0p7K} ($x=0.000$, $T=0.7$ K), and 
Fig.~\ref{QO_AMATERAS_Intensity_map} ($x=0.003$, $T=0.1$ K). 
Numerical uncertainty of $J_m$ is discussed in Appendix~\ref{appendix_LS} and Ref.~\cite{Supplemental_Material}.
}
\begin{ruledtabular}
\begin{tabular}{ccccccccccccccc}
3D data & $J_1$ & $J_2$ & $J_3$ & $J_4$ & $J_5$ & $J_6$ & $J_7$ & $J_8$ & $J_9$ & $J_{10}$ & $J_{11}$ & $J_{12}$ & $J_{13}$ & $J_{14}$\\ \hline
Fig.~\ref{QSL_IN5_Intensity_map0p1K} & 1.0 & 0.824 & 1.011 & 0.176 & 0.184 & 0.410 & 0.436 & 0.355 & 1.060 & -0.026 & -0.066 & -0.071 & 0.378 & \\
Fig.~\ref{QSL_IN5_Intensity_map0p7K} & 1.0 & 0.070 & 0.536 & -0.373 & -0.370 & 0.076 & -0.007 & -0.020 & 0.919 \\
Fig.~\ref{QO_IN5_Intensity_map0p1K} & 1.0 & 0.836 & 1.191 & 0.102 & 0.109 & 0.487 & 0.745 & 0.574 & 1.732 & 0.037 & 0.014 & -0.137 & 0.464 & \\
Fig.~\ref{QO_IN5_Intensity_map0p7K} & 1.0 & -0.101 & 0.751 & -0.501 & -0.408 & 0.191 & 0.078 & -0.019 & 1.364 \\
Fig.~\ref{QO_AMATERAS_Intensity_map} & 0.25 & -0.279 & -0.040 & -0.237 & -0.081 & -0.124 & 0.297 & 0.022 & 0.098 & -0.061 & -0.031 & -0.060 & -0.119 & 0.191 \\
\end{tabular}
\end{ruledtabular}
\end{table*}
\begin{figure*}[htb]
\centering
\includegraphics[width=18.0cm,clip]{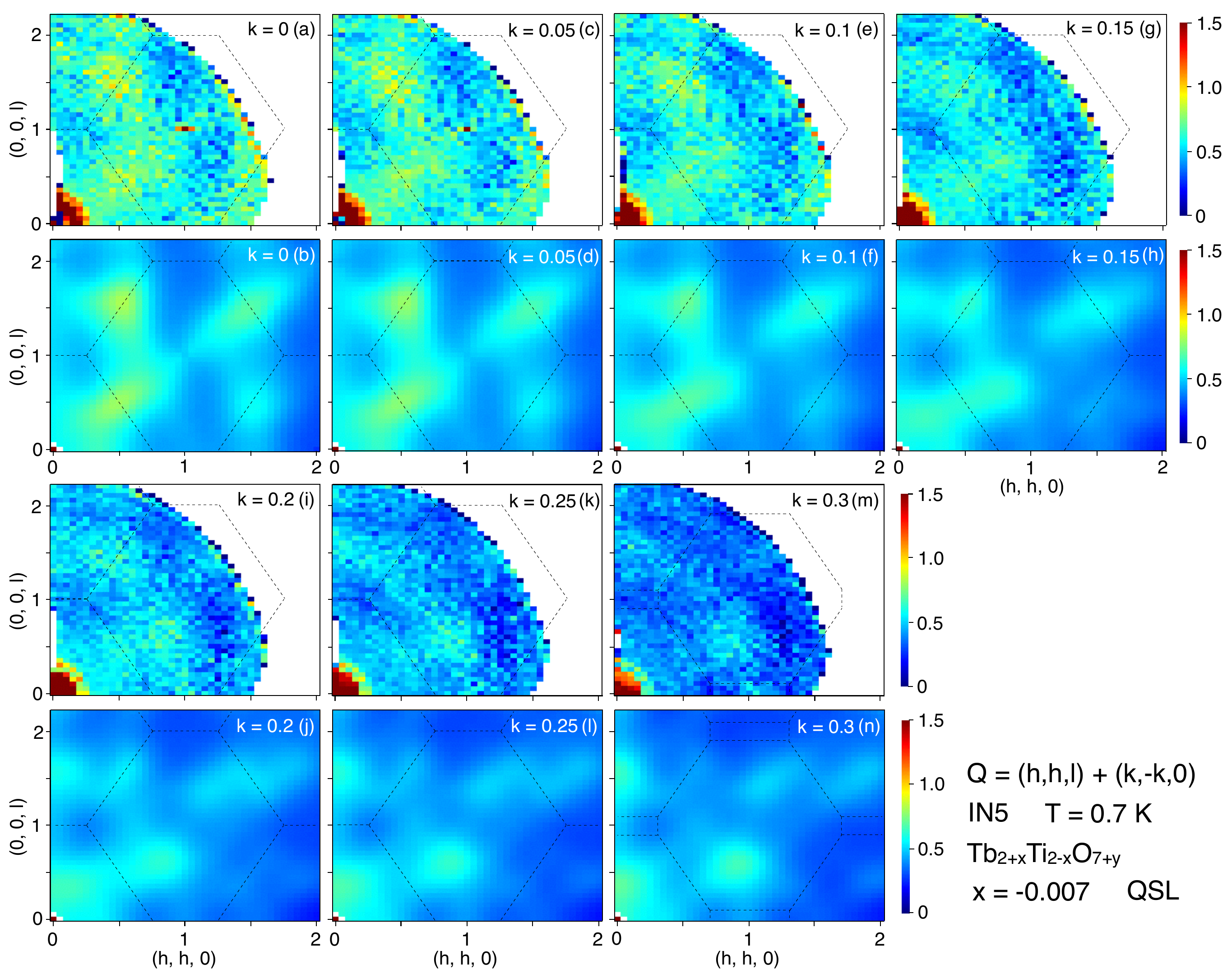}
\caption{ 
Intensity maps of 3D data $[S(\bm{Q})]_{\text{el}}$ 
taken at 0.7 K for the QSL sample with $x = -0.007$. 
The 3D data are viewed by 2D slices (a,c,e,g,i,k,m), which are parallel cross-sections 
of $\bm{Q}=(h,h,l)+(k,-k,0)$ with fixed $k$. 
These can be compared to the typical RPA $[S(\bm{Q})]_{\text{el}}$ (b,d,f,h,j,l,n) 
obtained by least squares fit 
using the 9 coupling constants, $J_1, \cdots, J_{9}$, 
listed in Table~\ref{FitJ1J14}. 
Dashed lines in these 2D slices (a-n) are boundaries of Brillouin zones. 
}
\label{QSL_IN5_Intensity_map0p7K}
\end{figure*}
Figure~\ref{QSL_IN5_Intensity_map0p1K}(a,c,e,g,i,k,m) shows 
a 3D data set $[S(\bm{Q})]_{\text{el}}$ 
taken at 0.1 K for the QSL sample with $x=-0.007$. 
These 3D data are shown by seven 2D slices 
of $\bm{Q}=(h,h,l)+(k,-k,0)$ with fixed $k$ values. 
Two slice planes with $k=0$ and 0.25 are 
illustrated at the bottom right corner of Fig.~\ref{QSL_IN5_Intensity_map0p1K} 
with the first Brillouin zone of the FCC lattice and an irreducible zone. 
From this figure one can see that the observed $\bm{Q}$-range 
encompasses an independent part of the first Brillouin zone, 
which is an advantage over the previous 
experiments, which is limited to the 2D slice with $k=0$ \cite{Yasui2002,Fennell2012,Petit12,Fritsch13}. 

The observed 3D data $[S(\bm{Q})]_{\text{el}}$ of Fig.~\ref{QSL_IN5_Intensity_map0p1K} 
show two features: 
strong short-ranged spin correlations with 
wavevector 
$\bm{q} = (\tfrac{1}{2},\tfrac{1}{2},\tfrac{1}{2})$, 
and very weak pinch-point structures around $\bm{Q}=(1,1,1)$ and 
$(0,0,2)$. 
By comparing the 2D slice of Fig.~\ref{QSL_IN5_Intensity_map0p1K}(a) 
with those of the previous investigations \cite{Yasui2002,Fennell2012,Petit12,Fritsch13}, 
one can see both differences and similarities 
among the investigations. 
This fact confirms the importance of controlling the $x$ value 
for quantitative studies. 

In order to measure the 
temperature dependence of 
the $\bm{q} = (\tfrac{1}{2},\tfrac{1}{2},\tfrac{1}{2})$ SRO 
we measured intensities along a trajectory 
through $\bm{Q} = (\tfrac{1}{2},\tfrac{1}{2},\tfrac{1}{2})$ 
by fixing the sample rotation angle. 
The resulting temperature dependence of $[S(\bm{Q})]_{\text{el}}$ 
is plotted in Fig.~\ref{QSL_HalfHalfHalf_Tdep}. 
As temperature is decreased below 0.4 K, 
the spin correlations grow continuously without a phase transition. 
We estimate the correlation length $\xi$ from 
the half width at half maximum (HWHM) of the peak ($1/\xi$ = HWHM). 
It increases to $\xi \sim 20$ {\AA} at 0.1 K. 
This correlation length and the temperature scale of 0.4 K 
agree with those reported in Ref.~\cite{Guitteny2015}, 
where powder samples were used (Fig.~3(b) in Ref.~\cite{Guitteny2015}). 
We note that 
the correlation length reported in Ref.~\cite{Fritsch13}, 
where a single crystal sample was used, is significantly shorter ($\sim 8$ {\AA}). 

An important point concerning 
the discrepancy of the correlation length 
noted above concerns the thermal response time of the system. 
In particular, we observed very slow cooling of the sample 
especially below 0.4 K in the present experimental condition. 
More specifically, it took about two days for the scattering intensity 
to become time independent after cooling the mixing chamber down to 0.1 K. 
This slow cooling is ascribable to very low thermal conductivity of TTO \cite{Li2013} 
and the large size of the crystal sample for INS. 
One has to carefully distinguish this long relaxation time 
to other interpretations, for example, 
the cooling protocol dependence reported in Ref.~\cite{Kermarrec2015}, 
where the authors might not have waited enough time, 
which may possibly result in a short correlation length. 

We performed least squares fits of the observed 3D data set $[S(\bm{Q})]_{\text{el}}$ 
to the RPA intensity Eq.~(\ref{integratedSQEelastic}). 
Adjustable parameters are the coupling constants 
$J_m$ ($1 \leq m \leq m_{\text{max}}$), 
the local susceptibility $\chi_{\text{L}}$, 
and an intensity scale factor. 
After several trial computations, 
we became aware of a problem that  
these parameters cannot be independently adjusted. 
To avoid this problem and exclude unrealistic solutions, 
we fixed $J_1$ and imposed a restriction on $J_m$ ($2 \leq m \leq m_{\text{max}}$) 
by adding a penalty function 
$\sum_{2 \leq m \leq m_{\text{max}}} \left( \tfrac{J_m}{1 \text{ K}} \right)^8$ 
to the weighted sum of squared residuals 
\begin{equation}
\chi^2 = \sum_{i=1}^N \left( \tfrac{\text{obs}(i) - \text{calc}(i)}{\text{error}(i)} \right)^2 \: ,
\label{chi2}
\end{equation}
where $N=10185$ is the number of intensity data used in the fitting. 
Technical details of the least squares fits are discussed 
in Appendix~\ref{appendix_LS} and Ref.~\cite{Supplemental_Material}.

In Fig.~\ref{chisq_QSL_IN5_0p1K0p7K}(a) we plot minimized values of $\chi^2$ 
as a function of fixed $J_1$ 
(detailed discussion on inspecting 
the least squares fits is given in Ref.~\cite{Supplemental_Material}). 
As $J_1$ is decreased in the range $J_1 < -5D/3$, 
which favors the antiferromagnetic ``all-in--all-out'' 
LRO for $J_{m>1} =0$ \cite{Hertog00}, 
the fits become unsatisfactory. 
These plots also show that 
the inclusion of further 
coupling constants $J_m$ with $m_{\text{max}} \ge 14$ 
does not improve the fitting. 

By inspecting 3D data $[S(\bm{Q})]_{\text{el}}$ calculated using several sets of fitted parameters, 
we chose a typical good result of the fitting.
This typical $[S(\bm{Q})]_{\text{el}}$ is shown in Fig.~\ref{QSL_IN5_Intensity_map0p1K}(b,d,f,h,j,l,n), 
which is calculated 
using the values of $J_1, \cdots , J_{13}$ listed in Table~\ref{FitJ1J14}. 
One can see that 
the RPA model calculation excellently reproduces the observed $[S(\bm{Q})]_{\text{el}}$. 
Almost the same features of the $\bm{q} = (\tfrac{1}{2},\tfrac{1}{2},\tfrac{1}{2})$ SRO, 
the very weak pinch point structures, 
and the other structures in $\bm{Q}$-space are seen in 
both the observed and calculated $[S(\bm{Q})]_{\text{el}}$. 
This goodness of fit indicates that the QSL sample retains 
the space group symmetry of the pyrochlore structure ($Fd\bar{3}m$) 
as low as 0.1 K. 
The coupling constants listed in Table~\ref{FitJ1J14} 
are much larger than 
those expected for bare exchange interactions; 
for example, the \nth{7} 
neighbor coupling $J_9$ is as large as the nearest neighbor $J_1$. 
This fact indicates either that the coupling constants are strongly renormalized, e.g., 
by integrating out excited states with $E> \epsilon$, 
or that the present analysis is an experimental parametrization. 

Figure~\ref{QSL_IN5_Intensity_map0p7K}(a,c,e,g,i,k,m) shows 
a 3D data set $[S(\bm{Q})]_{\text{el}}$ 
taken at 0.7 K for the QSL sample with $x=-0.007$. 
The image contrast of this $[S(\bm{Q})]_{\text{el}}$ becomes much lower than that of 0.1 K. 
Only a slight trace of 
the $\bm{q} = (\tfrac{1}{2},\tfrac{1}{2},\tfrac{1}{2})$ SRO is seen. 
On the other hand, quite intriguingly, 
the pinch point structure around $\bm{Q}=(1,1,1)$ 
becomes clearer and bears a resemblance to that observed for 
the spin ice compound Ho$_2$Ti$_2$O$_7$ \cite{Bramwell01,Fennell2009}. 
This agrees with our proposal \cite{Takatsu2016prl} 
that the magnetic part of the pseudospin-$\frac{1}{2}$ Hamiltonian of TTO 
is that of dipolar spin ice \cite{Hertog00}. 

We performed least squares fits of the observed 3D data set $[S(\bm{Q})]_{\text{el}}$ 
to the RPA intensity Eq.~(\ref{integratedSQEelastic}) 
in the same way as those of 0.1 K. 
In Fig.~\ref{chisq_QSL_IN5_0p1K0p7K}(b) we plot minimized values of $\chi^2$ 
as a function of the fixed $J_1$. 
This figure shows that as $J_1$ is decreased in the range $J_1 < -5D/3$, 
the fits become unsatisfactory, 
and that the inclusion of further 
coupling constants $J_m$ with $m_{\text{max}} \ge 10$ 
does not improve the fitting. 
By inspecting several calculated $[S(\bm{Q})]_{\text{el}}$, 
we chose a typical good result of the fitting. 
This typical $[S(\bm{Q})]_{\text{el}}$ is shown in Fig.~\ref{QSL_IN5_Intensity_map0p7K}(b,d,f,h,j,l,n), 
which is calculated 
using the values of $J_1, \cdots , J_{9}$ listed in Table~\ref{FitJ1J14}. 
Considering the lower image contrast and larger statistical errors, 
the agreement is acceptably good. 
In fact, both the weakly peaked structures with $\bm{q} = (\tfrac{1}{2},\tfrac{1}{2},\tfrac{1}{2})$ 
and the pinch point structure around $\bm{Q}=(1,1,1)$ 
are reproduced in the RPA $[S(\bm{Q})]_{\text{el}}$. 
It should be noted that 
the typical coupling constants listed in the first (0.1 K) and second (0.7 K) lines 
in Table~\ref{FitJ1J14} are considerably different. 
This strong temperature dependence also suggests that 
the fitted values of the coupling constants are 
either renormalized values or experimental parameters. 
We also note that at 0.7 K the largest $J_m$ is $J_1=1.0$ K, 
which favors the spin ice state and agrees with 
our estimation of $J_{nn}$ ($=J_1$) 
based on high temperature susceptibility ($T>5$ K) \cite{Takatsu2016prl}, 
which may possibly support the interpretation that $J_m$ are renormalized at low temperatures. 

\subsection{\label{results_IN5_QO} QO sample with $x=0.000$}
\begin{figure*}[htb]
\centering
\includegraphics[width=18.0cm,clip]{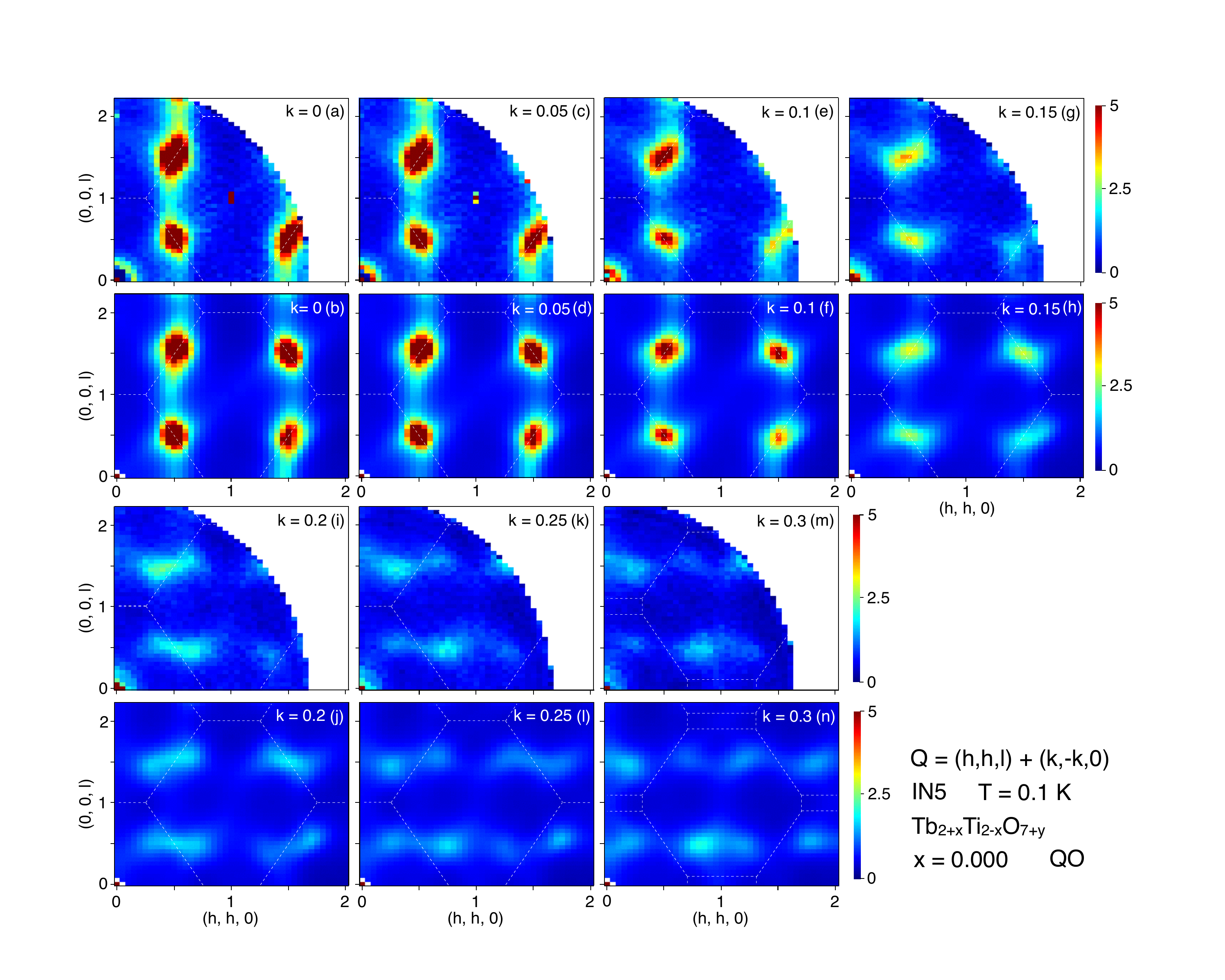}
\caption{ 
Intensity maps of 3D data $[S(\bm{Q})]_{\text{el}}$ 
taken at 0.1 K for the QO sample with $x = 0.000$. 
The 3D data are viewed by 2D slices (a,c,e,g,i,k,m), which are parallel cross-sections 
of $\bm{Q}=(h,h,l)+(k,-k,0)$ with fixed $k$. 
These can be compared to the typical RPA $[S(\bm{Q})]_{\text{el}}$ (b,d,f,h,j,l,n) 
obtained by least squares fit 
using the 13 coupling constants, $J_1, \cdots, J_{13}$, 
listed in Table~\ref{FitJ1J14}. 
Dashed lines in these 2D slices (a-n) are boundaries of Brillouin zones. 
}
\label{QO_IN5_Intensity_map0p1K}
\end{figure*}
\begin{figure*}[htb]
\centering
\includegraphics[width=18.0cm,clip]{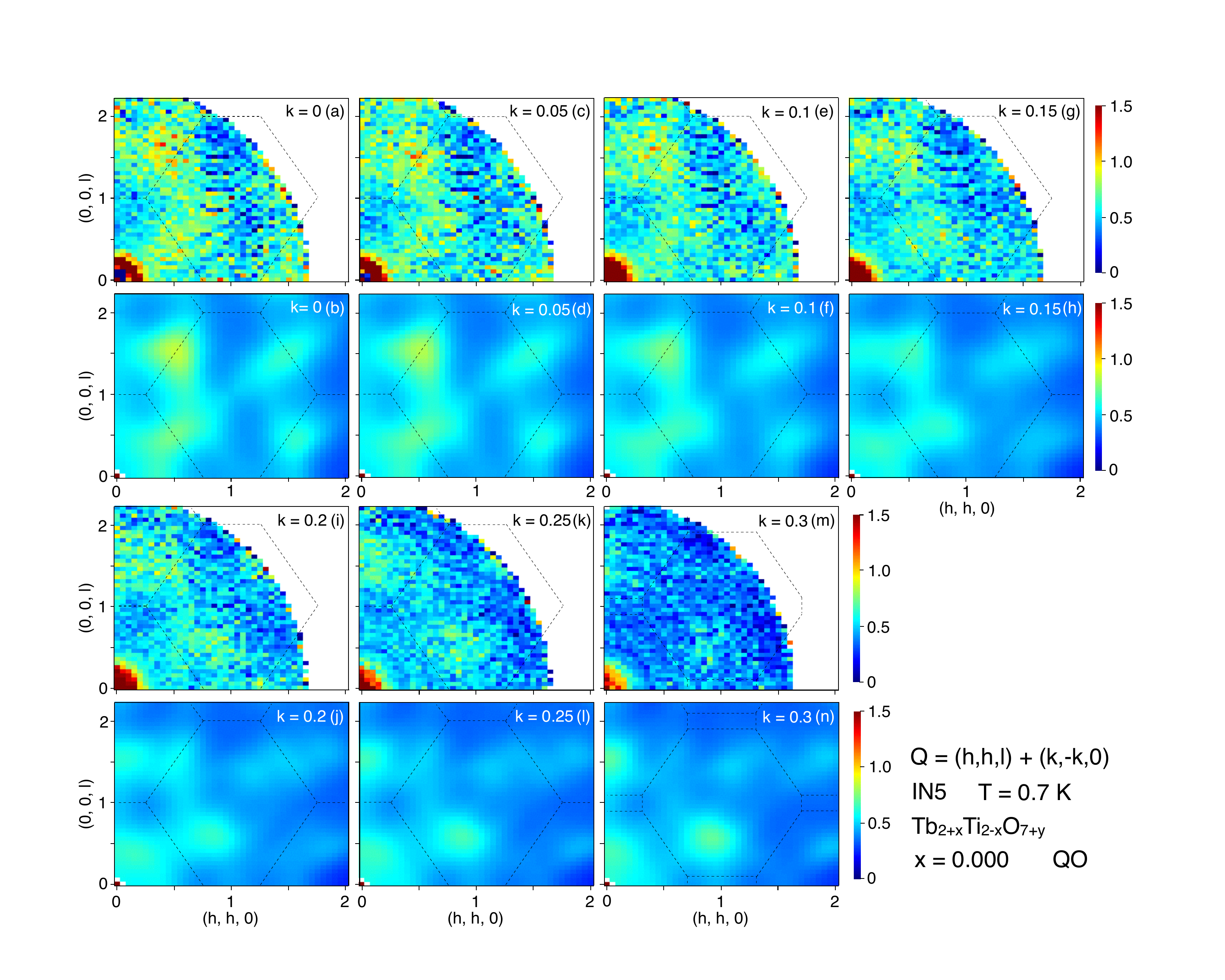}
\caption{ 
Intensity maps of 3D data $[S(\bm{Q})]_{\text{el}}$ 
taken at 0.7 K for the QO sample with $x = 0.000$. 
The 3D data are viewed by 2D slices (a,c,e,g,i,k,m), which are parallel cross-sections 
of $\bm{Q}=(h,h,l)+(k,-k,0)$ with fixed $k$. 
These can be compared to the typical RPA $[S(\bm{Q})]_{\text{el}}$ (b,d,f,h,j,l,n) 
obtained by least squares fit using the 9 coupling constants, $J_1, \cdots, J_{9}$, 
listed in Table~\ref{FitJ1J14}. 
Dashed lines in these 2D slices (a-n) are boundaries of Brillouin zones. 
}
\label{QO_IN5_Intensity_map0p7K}
\end{figure*}
\begin{figure}[htb]
\centering
\includegraphics[width=8.0cm,clip]{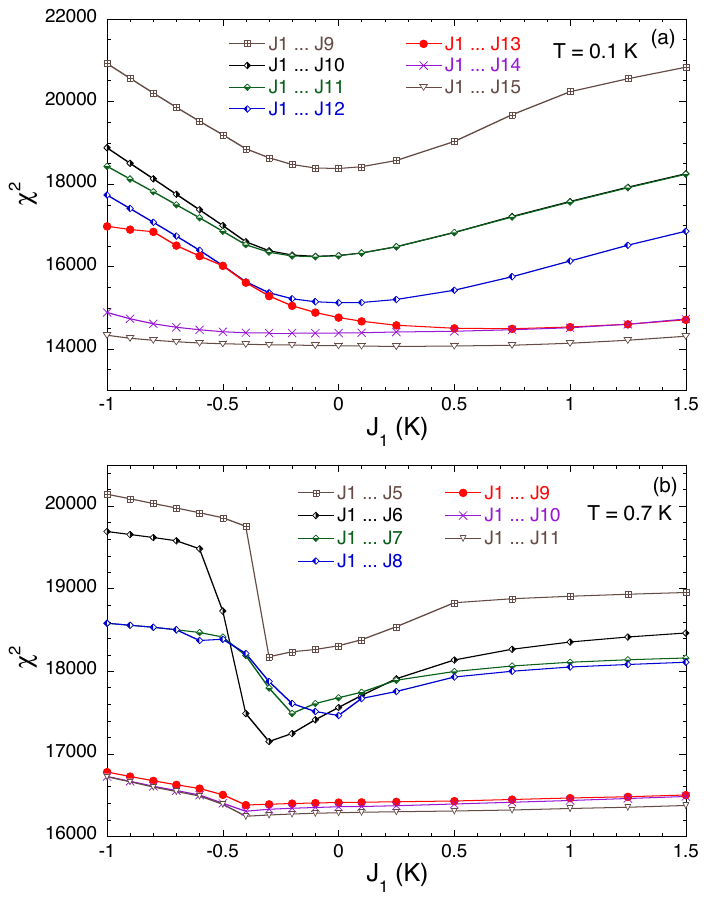}
\caption{ 
Minimized values of the weighted sum of squared residuals 
$\chi^2$ as a function of the fixed parameter $J_1$. 
(a) Results of  
least squares fits of $[S(\bm{Q})]_{\text{el}}$ with adjustable parameters 
$J_m$ ($m \le m_{\text{max}} = 9, \cdots , 15$)
for the QO sample with $x=0.000$ taken at $0.1$ K 
(Fig.~\ref{QO_IN5_Intensity_map0p1K}). 
The number of fit data is 11418. 
(b) Results of  
least squares fits of $[S(\bm{Q})]_{\text{el}}$ with adjustable parameters 
$J_m$ ($m \le m_{\text{max}} = 5, \cdots , 11$)
for the QO sample with $x=0.000$ taken at 0.7 K 
(Fig.~\ref{QO_IN5_Intensity_map0p7K}).
The number of fit data is 10520. 
}
\label{chisq_QO_IN5_0p1K0p7K}
\end{figure}
We show 3D data sets $[S(\bm{Q})]_{\text{el}}$ 
for the QO sample with $x=0.000$ 
taken at 0.1 and 0.7 K 
in Fig.~\ref{QO_IN5_Intensity_map0p1K}(a,c,e,g,i,k,m) 
and Fig.~\ref{QO_IN5_Intensity_map0p7K}(a,c,e,g,i,k,m), 
respectively. 
By comparing these figures with the corresponding $[S(\bm{Q})]_{\text{el}}$ shown in 
Fig.~\ref{QSL_IN5_Intensity_map0p1K} and Fig.~\ref{QSL_IN5_Intensity_map0p7K} 
for the QSL sample, 
one can see that the 3D data $[S(\bm{Q})]_{\text{el}}$ of these QSL and QO samples 
show many similarities, which suggests a common origin. 
This is in stark contrast to the difference of their inelastic spectra 
shown in Fig.~2 of Ref.~\cite{Kadowaki2018}. 
Close inspection of the 3D data $[S(\bm{Q})]_{\text{el}}$ of Fig.~\ref{QO_IN5_Intensity_map0p1K} 
and Fig.~\ref{QSL_IN5_Intensity_map0p1K} 
shows that the peaked structures at 
$\bm{Q} = (\tfrac{1}{2},\tfrac{1}{2},\tfrac{1}{2})$ 
and $(\tfrac{1}{2},\tfrac{1}{2},\tfrac{3}{2})$ 
of the QO sample are slightly broader than those of the QSL sample, 
and that the peak width of the QO sample is slightly larger than the QSL sample. 
This indicates that the small quadrupole order 
slightly suppresses the $\bm{q} = (\tfrac{1}{2},\tfrac{1}{2},\tfrac{1}{2})$ SRO. 

We performed least squares fits of the observed 3D data sets $[S(\bm{Q})]_{\text{el}}$ 
to the RPA intensity Eq.~(\ref{integratedSQEelastic}), 
in the same way as those of the QSL sample. 
Resulting minimized values of $\chi^2$ are plotted 
as a function of the fixed $J_1$ 
in Fig.~\ref{chisq_QO_IN5_0p1K0p7K}(a) and (b) 
for the 0.1 and 0.7 K data, respectively. 
These figures and Figs.~\ref{chisq_QSL_IN5_0p1K0p7K}(a) and (b) show 
that the least squares fits provided parallel results with those 
of the QSL sample. 
In fact, the typical coupling constants obtained by the fits, 
which are listed in Table~\ref{FitJ1J14}, 
have many similarities for the two samples both at 0.1 and 0.7 K. 
Using these typical $J_m$ listed in Table~\ref{FitJ1J14} 
we calculated RPA $[S(\bm{Q})]_{\text{el}}$ and show them 
in Fig.~\ref{QO_IN5_Intensity_map0p1K}(b,d,f,h,j,l,n) 
and Fig.~\ref{QO_IN5_Intensity_map0p7K}(b,d,f,h,j,l,n). 
The observed and the calculated $[S(\bm{Q})]_{\text{el}}$ 
agree excellently and acceptably well at 0.1 K and 0.7 K, respectively. 

\subsection{\label{results_AMATERAS} QO sample with $x=0.003$}
\begin{figure*}[htb]
\centering
\includegraphics[width=18.0cm,clip]{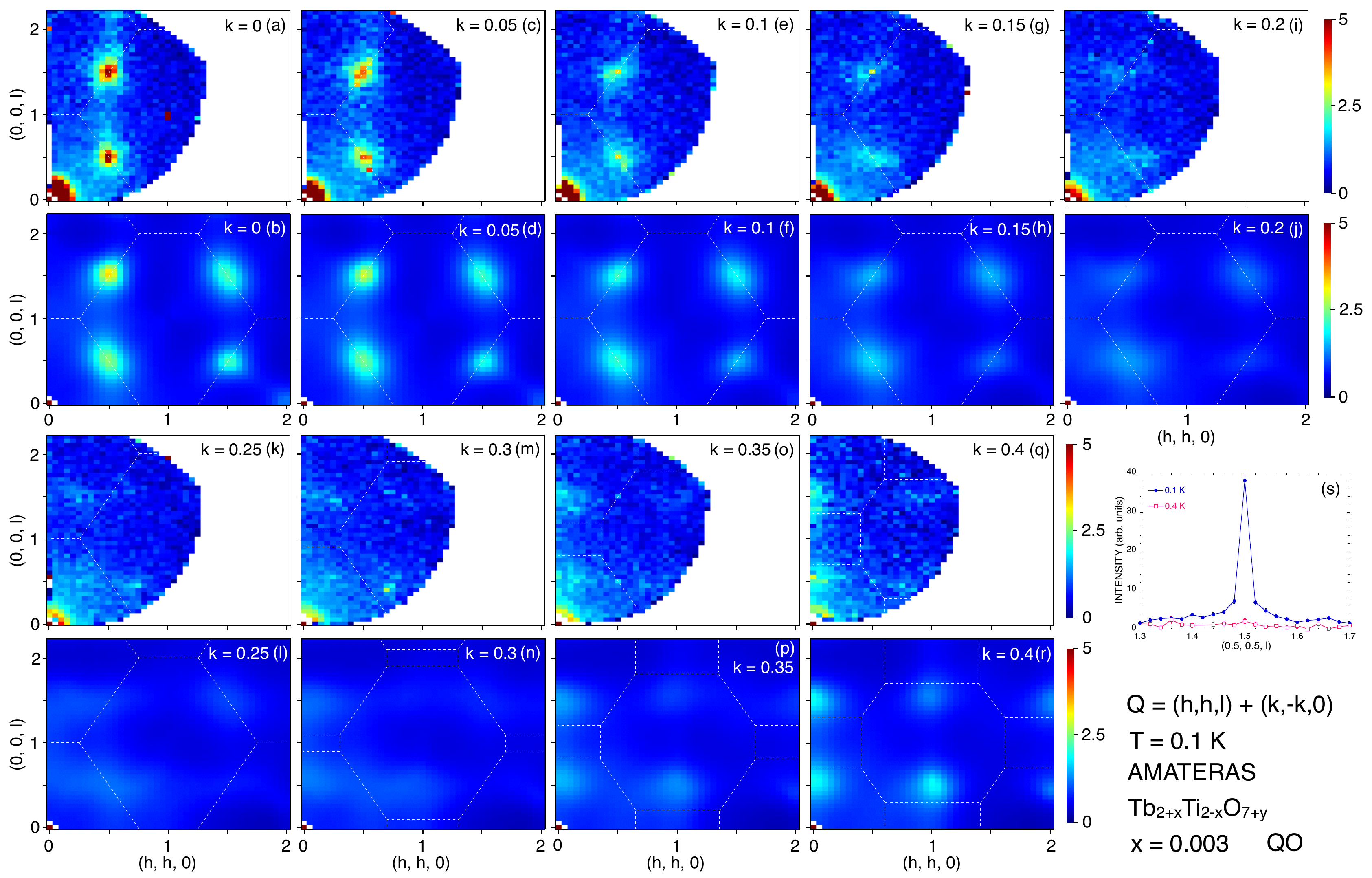}
\caption{ 
Intensity maps of 3D data $[S(\bm{Q})]_{\text{el}}$ 
taken at 0.1 K for the QO sample with $x = 0.003$. 
The 3D data are viewed by 2D slices (a,c,e,g,i,k,m,o,q), which are parallel cross-sections 
of $\bm{Q}=(h,h,l)+(k,-k,0)$ with fixed $k$. 
These can be compared to the typical RPA $[S(\bm{Q})]_{\text{el}}$ (b,d,f,h,j,l,n,p,r) 
obtained by least squares fit 
using the 14 coupling constants, $J_1, \cdots, J_{14}$, 
listed in Table~\ref{FitJ1J14}. 
Dashed lines in these 2D slices (a-r) are boundaries of Brillouin zones. 
(s) $Q$-scan along $\bm{Q}=(\tfrac{1}{2},\tfrac{1}{2},l)$ close to magnetic reflection 
$(\tfrac{1}{2},\tfrac{1}{2},\tfrac{3}{2})$. 
}
\label{QO_AMATERAS_Intensity_map}
\end{figure*}
\begin{figure}[htb]
\centering
\includegraphics[width=8.0cm,clip]{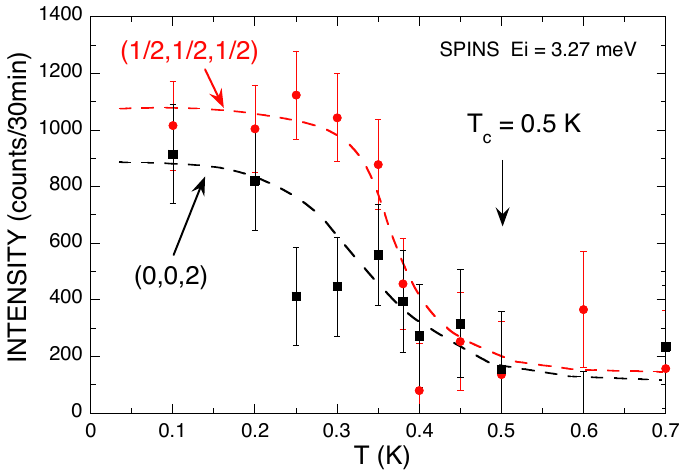}
\caption{ 
Temperature dependence of intensity of Bragg reflection 
of the powder sample with $x=0.005$ used in Ref.~\cite{Taniguchi13}. 
These data were measured on the triple-axis spectrometer SPINS operated with $\lambda = 5$ {\AA} at NIST. 
Error bars represent one standard deviation. 
}
\label{QO_0p005_Bragg}
\end{figure}
\begin{figure}[htb]
\centering
\includegraphics[width=8.0cm,clip]{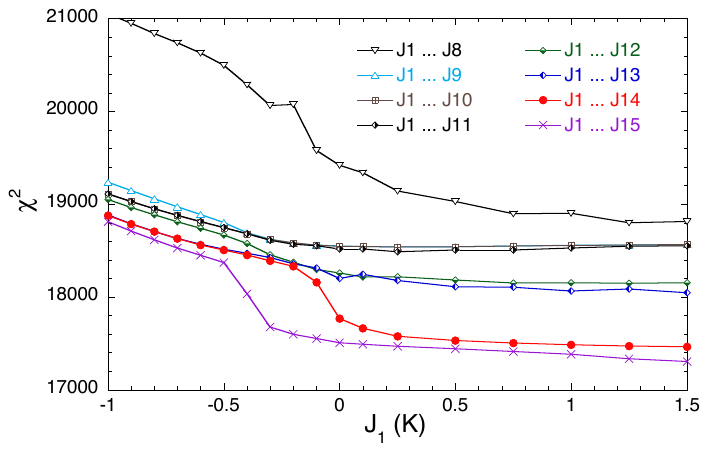}
\caption{ 
Minimized values of the weighted sum of squared residuals 
$\chi^2$ are plotted as a function of the fixed parameter $J_1$. 
These are obtained by 
least squares fits of $[S(\bm{Q})]_{\text{el}}$ with adjustable parameters 
$J_m$ ($m \le m_{\text{max}} = 8, \cdots , 15$) 
for the QO sample with $x=0.003$ taken at $0.1$ K 
(Fig.~\ref{QO_AMATERAS_Intensity_map}). 
The number of fit data is 10570. 
}
\label{chisq_QO_AMATERAS}
\end{figure}
\begin{figure*}[htb]
\centering
\includegraphics[width=18.0cm,clip]{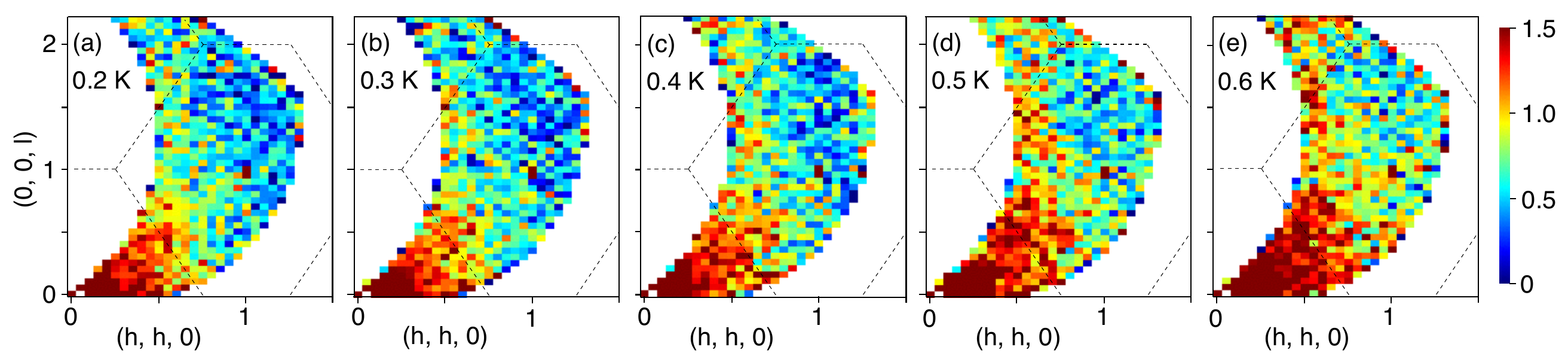}
\caption{ 
Temperature dependence of 
$[S(\bm{Q})]_{\text{el}}$ in the $(h,h,l)$ plane, where $k=0$ is fixed, 
for the QO sample with $x = 0.003$. 
Dashed lines are boundaries of Brillouin zones. 
}
\label{QO_AMATERAS_2D_map_Tdep}
\end{figure*}
Figure~\ref{QO_AMATERAS_Intensity_map}(a,c,e,g,i,k,m,o,q) shows 
a 3D data set $[S(\bm{Q})]_{\text{el}}$ 
taken at 0.1 K for the QO sample with $x=0.003$. 
These 3D data are substantially different from those 
of the QSL sample and the QO sample with $x=0.000$. 
The pinch point structure disappears. 
The $\bm{q}=(\tfrac{1}{2},\tfrac{1}{2},\tfrac{1}{2})$ SRO becomes 
much broader than that of the QO sample with $x=0.000$. 
Another new point of this sample is that there appears a tiny magnetic Bragg reflection 
at $\bm{Q} = (\tfrac{1}{2},\tfrac{1}{2},\tfrac{3}{2})$.
A $Q$-scan through this reflection is plotted in Fig.~\ref{QO_AMATERAS_Intensity_map}(s), 
which shows that it disappears at 0.4 K. 
We note that detector gaps of AMATERAS prohibited us from measuring 
$\bm{Q} = (\tfrac{1}{2},\tfrac{1}{2},\tfrac{1}{2})$ and $(0,0,2)$ 
reflections. 

The appearance of tiny magnetic Bragg reflections at 
$\bm{Q} = (\tfrac{1}{2},\tfrac{1}{2},\tfrac{3}{2})$, 
$(\tfrac{1}{2},\tfrac{1}{2},\tfrac{1}{2})$, and 
$(0,0,2)$ was reported only for samples with 
large quadrupole orders \cite{Taniguchi13,Takatsu2016prl,Guitteny2015}. 
In order to complement our previous experimental data of the magnetic Bragg reflections 
shown in Fig.~5 of Ref.~\cite{Taniguchi13}, 
we show temperature dependence of intensities of the Bragg reflections 
at $\bm{Q} = (\tfrac{1}{2},\tfrac{1}{2},\tfrac{1}{2})$ and $(0,0,2)$ 
in Fig.~\ref{QO_0p005_Bragg}. 
Although statistical errors are large, one can see that the temperature dependence 
agrees with that shown in Fig.~3 of Ref.~\cite{Takatsu2016prl}. 
Since several observations of the magnetic Bragg reflections have been accumulated, 
one may now have to accept the conclusion that 
the tiny magnetic Bragg reflections, 
indicating LRO of magnetic moments of the order $\sim 0.1 \mu_{\text{B}}$, 
have a common origin attributed to the quadrupole LRO. 
They may possibly be caused by multi-spin interactions \cite{Molavian2009,Rau_Gingras2018}, 
which couple the magnetic and quadrupole moments. 

We performed least squares fits of the 3D data set $[S(\bm{Q})]_{\text{el}}$ 
to the RPA intensity Eq.~(\ref{integratedSQEelastic}), 
in the same way as the QSL sample. 
In Fig.~\ref{chisq_QO_AMATERAS} we plot minimized values of $\chi^2$ 
as a function of the fixed $J_1$. 
This figure shows that as $J_1$ is decreased in the range $J_1 < -5D/3$, 
the fits become unsatisfactory, 
and that the inclusion of further 
coupling constants $J_m$ with $m_{\text{max}} \ge 15$ does not improve the fitting. 
By inspecting several calculated $[S(\bm{Q})]_{\text{el}}$, 
we chose a typical good result of the fitting.
This typical $[S(\bm{Q})]_{\text{el}}$ is shown in Fig.~\ref{QO_AMATERAS_Intensity_map}(b,d,f,h,j,l,n,p,r), 
which is calculated using the values of $J_1, \cdots , J_{14}$ listed in Table~\ref{FitJ1J14}. 
One can see that the agreement between 
the calculated and observed $[S(\bm{Q})]_{\text{el}}$ 
is not as good as that of the QSL sample. 
This less satisfactory agreement suggests that the quadrupole order breaks 
the space group symmetry. 
In fact, the proposed quadrupole order in Ref.~\cite{Takatsu2016prl} breaks this symmetry.
We note that the typical coupling constants obtained by the fitting (Table~\ref{FitJ1J14}) 
are substantially different from those of the QSL sample. 

Figure \ref{QO_AMATERAS_2D_map_Tdep} shows the temperature dependence 
of 2D intensity map in the plane $\bm{Q}=(h,h,l)$ 
observed in a temperature range $0.2 \le T \le 0.6$ K. 
Although the $\bm{Q}$ range and statistical errors are limited, 
these 2D maps show that 
the $\bm{q} = (\tfrac{1}{2},\tfrac{1}{2},\tfrac{1}{2})$ SRO 
disappears already at 0.2 K. 
The pinch point structure around $(1,1,1)$, 
which is similar to that of the QSL sample at 0.7 K, 
is barely observable in the 0.3 and 0.4 K data. 
In the temperature range above 0.5 K, 
where the electric quadrupole order disappears,
another kind of spin correlations seems to develop. 

\section{Discussion}
A question of ``what does $[S(\bm{Q})]_{\text{el}}$ measure?'' 
is a little difficult to answer correctly. 
By the present definition, 
the (nominally) elastic scattering intensity 
$[S(\bm{Q})]_{\text{el}} = \int_{-\epsilon}^{\epsilon} S(\bm{Q},E) dE $ 
is defined on the basis of the present experimental conditions; 
thereby $[S(\bm{Q})]_{\text{el}}$ is different from 
theoretically elastic scattering. 
For the sake of simplicity as well as for our interest in the QLS state, 
we would like to discuss $[S(\bm{Q})]_{\text{el}}$ 
at the lowest temperature of the present experiments ($T = 0.1$ K). 
Considering that this temperature scale is approximately equal to 
the instrumental energy resolution scales, 
$[S(\bm{Q})]_{\text{el}}$ at 0.1 K is essentially (and roughly) expressed by 
\begin{equation}
\sum_{|E_i - E_{\text{G}}|,|E_j - E_{\text{G}}| < 0.1 \text{K}} 
\frac{e^{- \beta E_i}}{Z}
| \langle j | \sum_{\bm{r}} \sigma_{\bm{r}}^z e^{i \bm{Q} \cdot \bm{r} } | i \rangle |^2 \;, 
\label{SRO_el}
\end{equation}
where $E_{\text{G}}$ denotes the ground state energy 
and the summation runs over low-energy states, $| i \rangle$ and $| j \rangle$. 

In the previous analyses of 
the $\bm{q} = (\tfrac{1}{2},\tfrac{1}{2},\tfrac{1}{2})$ SRO \cite{Fritsch13,Guitteny2015}, 
a few static Ising-spin clusters were assumed to exist, 
where certain disorders suppressing LRO are also assumed implicitly. 
These assumptions would be justified, 
if the system behaved within the classical spin paradigm, 
where the states $| i \rangle$ and $| j \rangle$ in Eq.~(\ref{SRO_el}) are 
expressed simply by single states described by the Ising-spin clusters. 
However, when quantum effects are included 
the simple low-energy states would be replaced by 
linear combinations of the Ising-spin-cluster states. 
As the number of Ising-spin-cluster states 
in a linear combination is increased, 
the system departs from the classical spin paradigm, 
and consequently 
the cluster analyses \cite{Fritsch13,Guitteny2015} would not work properly. 
We speculate that the failures of obtaining sufficient goodness-of-fit 
in Refs.~\cite{Fritsch13,Guitteny2015} indicates that this really happened. 
For the present RPA analyses, 
although RPA takes account of quantum effects 
to a certain extent, 
RPA is basically a classical approach and thereby the same problem would occur, 
especially when quantum effects become substantially large, e.g., QSL states. 
We speculate that the breakdown of the classical paradigm 
is manifested as the necessity of the unexpectedly large number of 
coupling constants in the present RPA fitting. 

The observed $[S(\bm{Q})]_{\text{el}}$ shown in Fig.~\ref{QSL_IN5_Intensity_map0p1K} 
can be excellently reproduced 
by the RPA formulae Eq.~(\ref{GsusE0}) and Eq.~(\ref{integratedSQEelastic}). 
We think that there are two reasons for this successful fit. 
Firstly, the RPA formulae act as inverse Fourier transform. 
The many coupling constants imply 
that many inverse Fourier components are needed to 
reproduce the observed $[S(\bm{Q})]_{\text{el}}$. 
For example, the terms related to $J_3$ ($> 0$) in Eq.~(\ref{GsusE0}) give rise to 
higher $[S(\bm{Q})]_{\text{el}}$ 
at wavevectors $\bm{Q}=(\tfrac{1}{2},\tfrac{1}{2},\tfrac{1}{2})$, 
$(\tfrac{1}{2},\tfrac{1}{2},\tfrac{3}{2})$ etc. 
Secondly, the coupling constants $J_m$ in Eq.~(\ref{Hmag}) 
are allowed by the space group symmetry.
As a consequence the RPA intensity formulae 
reflect the symmetry of the pyrochlore structure. 
In this sense, we may conclude that the QSL state of TTO retains the space group symmetry.

Apart from the analyses, 
one can obtain a few hints for further investigations of the QSL state of TTO directly 
from a few experimental facts. 
As discussed in section \ref{results_IN5_QSL}, 
the 3D data set $[S(\bm{Q})]_{\text{el}}$ at 0.7 K (Fig.~\ref{QSL_IN5_Intensity_map0p7K}) 
shows the pinch point structure around $\bm{Q}=(1,1,1)$. 
This suggests that the QSI state proposed in Ref.~\cite{Molavian07} 
is somehow continuously connected to the QSL state of TTO. 
The tiny magnetic Bragg reflections observed in several QO samples, 
discussed in section \ref{results_AMATERAS},  
are now regarded as an experimental fact. 
Thus the pseudospin-$\frac{1}{2}$ Hamiltonian is to be modified to include 
coupling between magnetic and quadrupole moments. 

\section{Conclusions}
Spin correlations of the frustrated pyrochlore oxide Tb$_{2+x}$Ti$_{2-x}$O$_{7+y}$ 
have been investigated by inelastic neutron scattering 
using single crystalline samples showing both the quantum-spin-liquid 
and quadrupole-ordered states.
The observed spin correlations show 
pinch-point type structures around $\Gamma$ points, 
an antiferromagnetic short-range order around $L$ points, 
and tiny antiferromagnetic Bragg scattering at $L$ and $\Gamma$ points. 
The $\bm{q}=(\tfrac{1}{2},\tfrac{1}{2},\tfrac{1}{2})$ short-range order was analyzed using a model calculation 
of a random phase approximation assuming two-spin interactions among Ising spins. 
Analyses have shown that the RPA scattering intensity 
well reproduces the experimental data 
using temperature and 
$x$ dependent coupling constants of up to \nth{10} neighbor site pairs. 
The unexpectedly large number of 
coupling constants required in the fitting suggest 
a breakdown of the classical spin paradigm at low temperatures 
and the necessity of a quantum spin paradigm. 
\begin{acknowledgments}
This work was supported by JSPS KAKENHI grant number 25400345. 
The neutron scattering performed using ILL IN5 (France) 
was transferred from JRR-3M HER (proposal 11567, 15545) 
with the approval of ISSP, Univ. of Tokyo, and JAEA, Tokai, Japan. 
The neutron scattering experiments at J-PARC AMATERAS were carried out 
under a research project number 2016A0327. 
The computation was performed on the CX400 supercomputer 
at the Information Technology Center, Nagoya University 
\cite{NISTdisclaimer}. 
\end{acknowledgments}

\appendix
\section{\label{appendix_RPA} RPA model calculation and definitions}

Methods of the RPA model calculation and related definitions 
are summarized in this section. 
The effective pseudospin-$\frac{1}{2}$ operators $\sigma_{\bm{r}}^{z}$ 
reside on the pyrochlore lattice sites $\bm{r}=\bm{t}_{n}+\bm{d}_{\nu}$, 
where $\bm{t}_n$ are FCC translation vectors and 
$\bm{d}_{\nu}$ are four crystallographic sites in the unit cell. 
These sites and their symmetry axes 
$\bm{x}_{\nu}$, $\bm{y}_{\nu}$, and $\bm{z}_{\nu}$ \cite{Kadowaki2015}
are listed in Table~\ref{local_axis}. 
Representative site pairs $\langle {\bm r} , {\bm r}^{\prime} \rangle_m $ 
of the coupling constants $J_m$ of Eq.~(\ref{Hmag}) 
are listed in Table~\ref{table_J1J16}.
\begin{table}
\caption{\label{local_axis}
Four crystallographic sites $\bm{d}_{\nu}$ ($\nu=0,1,2,3$) and their 
local symmetry axes $\bm{x}_{\nu}$, $\bm{y}_{\nu}$, and $\bm{z}_{\nu}$ \cite{Kadowaki2015}. 
}
\begin{ruledtabular}
\begin{tabular}{ccccc}
$\nu$ & $\bm{d}_{\nu}$ & $\bm{x}_{\nu}$ & $\bm{y}_{\nu}$ & $\bm{z}_{\nu}$ \\ \hline
0 & $\tfrac{1}{4}(0,0,0)$ & $\tfrac{1}{\sqrt{6}}(1,1,-2)$ & $\tfrac{1}{\sqrt{2}}(-1,1,0)$ & $\tfrac{1}{\sqrt{3}}(1,1,1)$  \\
1 & $\tfrac{1}{4}(0,1,1)$ & $\tfrac{1}{\sqrt{6}}(1,-1,2)$ & $\tfrac{1}{\sqrt{2}}(-1,-1,0)$ & $\tfrac{1}{\sqrt{3}}(1,-1,-1)$  \\
2 & $\tfrac{1}{4}(1,0,1)$ & $\tfrac{1}{\sqrt{6}}(-1,1,2)$ & $\tfrac{1}{\sqrt{2}}(1,1,0)$ & $\tfrac{1}{\sqrt{3}}(-1,1,-1)$  \\
3 & $\tfrac{1}{4}(1,1,0)$ & $\tfrac{1}{\sqrt{6}}(-1,-1,-2)$ & $\tfrac{1}{\sqrt{2}}(1,-1,0)$ & $\tfrac{1}{\sqrt{3}}(-1,-1,1)$  \\
\end{tabular}
\end{ruledtabular}
\end{table}
\begin{table}
\caption{\label{table_J1J16}
Representative site pairs 
$\langle {\bm r} , {\bm r}^{\prime} \rangle_m = \langle \bm{t}_{n}+\bm{d}_{\nu} , \bm{t}_{n^{\prime}}+\bm{d}_{\nu^{\prime}} \rangle_m$ 
of the coupling constants $J_m$ of Eq.~(\ref{Hmag}) 
are listed using $(\nu,\nu^{\prime})$ and 
$ {\bm r}^{\prime} - {\bm r} $. 
Distances between the site pairs $|{\bm r}^{\prime} - {\bm r}|$ 
show that the constants $J_m$ in this list are up to \nth{11} 
neighbor coupling, and that for \nth{3}, \nth{7}, \nth{9}, and \nth{10} 
neighbor site pairs, 
there are 2, 2, 3, and 2 non-equivalent site pairs, respectively. 
}
\begin{ruledtabular}
\begin{tabular}{cccc}
$J_m$ & $(\nu,\nu^{\prime})$ & ${\bm r}^{\prime} - {\bm r}$ & $|{\bm r}^{\prime} - {\bm r}|$ \\ \hline
$J_{1}$  & (0,1) &  (0   ,  1/4 ,  1/4) & 0.35355  \\  %1
$J_{2}$  & (0,1) &  (1/2 ,  1/4 , -1/4) & 0.61237  \\  %2
$J_{3}$  & (0,0) &  (1/2 ,  1/2 ,  0  ) & 0.70710  \\  %3 a
$J_{4}$  & (0,0) &  (1/2 , -1/2 ,  0  ) & 0.70710  \\  %3 b
$J_{5}$  & (0,1) &  (0   ,  3/4 , -1/4) & 0.79057  \\  %4
$J_{6}$  & (0,1) &  (1/2 ,  1/4 ,  3/4) & 0.93541  \\  %5
$J_{7}$  & (0,0) &  (1   ,  0   ,  0  ) & 1        \\  %6
$J_{8}$  & (0,1) &  (1   ,  1/4 ,  1/4) & 1.06066  \\  %7 a
$J_{9}$  & (0,1) &  (0   ,  3/4 ,  3/4) & 1.06066  \\  %7 b
$J_{10}$ & (0,1) &  (1/2 ,  3/4 , -3/4) & 1.17260  \\  %8
$J_{11}$ & (0,0) &  (1   , -1/2 , -1/2) & 1.224745 \\  %9 a
$J_{12}$ & (0,0) &  (1   ,  1/2 , -1/2) & 1.224745 \\  %9 b
$J_{13}$ & (0,0) &  (1   ,  1/2 ,  1/2) & 1.224745 \\  %9 c
$J_{14}$ & (0,1) &  (0   ,  5/4 ,  1/4) & 1.274755 \\  %10 a
$J_{15}$ & (0,1) &  (1   ,  3/4 , -1/4) & 1.274755 \\  %10 b
$J_{16}$ & (0,1) &  (1/2 ,  5/4 , -1/4) & 1.36930  \\  %11
\end{tabular}
\end{ruledtabular}
\end{table}

The generalized susceptibility $\chi_{\nu^{\prime \prime},\nu^{\prime}}(\bm{k}, E)$, 
where $\bm{k}$ is a vector in the FCC first Brillouin zone, 
is computed by solving an RPA equation \cite{Jensen91} 
\begin{equation}
\sum_{\nu^{\prime \prime}}[\delta_{\nu,\nu^{\prime \prime}} - \sum_{\nu^{\prime \prime \prime}}\chi^{0}_{\nu,\nu^{\prime \prime \prime}}(E) J_{\nu^{\prime \prime \prime},\nu^{\prime \prime}}(\bm{k})] \chi_{\nu^{\prime \prime},\nu^{\prime}}(\bm{k}, E) = \chi^{0}_{\nu,\nu^{\prime}}(E) \;,
\label{Gsus}
\end{equation}
where $J_{\nu,\nu^{\prime}}( \bm{k} )$ denotes 
the Fourier transform of the magnetic coupling constants 
$J_{n,\nu;n^{\prime},\nu^{\prime}}$ 
between sites $\bm{t}_{n^{\prime}}+\bm{d}_{\nu^{\prime}}$ and $\bm{t}_{n}+\bm{d}_{\nu}$ 
\begin{equation}
J_{\nu,\nu^{\prime}}(\bm{k}) = 
\sum_{n} J_{n,\nu;n^{\prime},\nu^{\prime}} 
e^{i \bm{k} \cdot [ (\bm{t}_{n}+\bm{d}_{\nu}) - (\bm{t}_{n^{\prime}}+\bm{d}_{\nu^{\prime}}) ]} \;,
\label{Ftransform}
\end{equation}
and $\chi^{0}_{\nu,\nu^{\prime}}(E)$ is the single site susceptibility. 
In the paramagnetic phase 
\begin{equation}
\chi^{0}_{\nu,\nu^{\prime}}(E) = 
\delta_{\nu,\nu^{\prime}} \chi_{\text{L}} \frac{\Gamma_0}{\Gamma_0 - i E} \;,
\label{sus0}
\end{equation}
where $\chi_{\text{L}}=1/(4 k_{\text{B}} T)$ is the local susceptibility \cite{Jensen91} 
and $\Gamma_0$ is a small positive constant. 

The neutron magnetic scattering intensity $S(\bm{Q}=\bm{G}+\bm{k},E)$, 
where $\bm{G}$ is a reciprocal lattice vector, 
is given by 
\begin{eqnarray}
& & S(\bm{Q},E) \propto f(Q)^{2} \frac{1}{1 - e^{-\beta E}} 
\sum_{\rho,\sigma,\nu,\nu^{\prime}} (\delta_{\rho,\sigma} - \hat{Q}_{\rho} \hat{Q}_{\sigma} ) \nonumber \\
& & \times  U_{\rho,z}^{(\nu)} U_{\sigma,z}^{(\nu^{\prime})} 
\text{Im} \left\{ \chi_{\nu,\nu^{\prime}}(\bm{k}, E) 
e^{- i \bm{G} \cdot (\bm{d}_{\nu} - \bm{d}_{\nu^{\prime}}) } \right\} \;,
\label{SQE}
\end{eqnarray}
where $U_{\rho,\alpha}^{(\nu)}$ is the rotation matrix
from the local ($\alpha$) frame defined at the sites 
$\bm{t}_n + \bm{d}_{\nu}$ to the global ($\rho$) frame \cite{Kao03,Kadowaki2015}. 
In the quasi-elastic approximation, 
the elastic scattering intensity $[S(\bm{Q})]_{\text{el}}$ is given by 
integrating Eq.~(\ref{SQE}) in a small range $|E|<\epsilon $ 
\begin{eqnarray}
& & [S(\bm{Q})]_{\text{el}} = \int_{-\epsilon}^{\epsilon} S(\bm{Q},E) dE \propto f(Q)^{2} 
\sum_{\rho,\sigma,\nu,\nu^{\prime}} (\delta_{\rho,\sigma} - \hat{Q}_{\rho} \hat{Q}_{\sigma} ) \nonumber \\
& & \times U_{\rho,z}^{(\nu)} U_{\sigma,z}^{(\nu^{\prime})} \int_{-\epsilon}^{\epsilon} 
\frac{ \text{Im} \left\{ \chi_{\nu,\nu^{\prime}}(\bm{k}, E) 
e^{- i \bm{G} \cdot (\bm{d}_{\nu} - \bm{d}_{\nu^{\prime}}) } \right\} }{E} dE \nonumber \\
& & \propto f(Q)^{2} \sum_{\rho,\sigma,\nu,\nu^{\prime}} (\delta_{\rho,\sigma} - \hat{Q}_{\rho} \hat{Q}_{\sigma} ) 
U_{\rho,z}^{(\nu)} U_{\sigma,z}^{(\nu^{\prime})} \nonumber \\
& & \;\;\;\;\;\;\;\;\;\;\;\; \times \chi_{\nu,\nu^{\prime}}(\bm{k}, 0) \cos [ \bm{G} \cdot (\bm{d}_{\nu} - \bm{d}_{\nu^{\prime}}) ] \; ,
\label{SQ_QEA}
\end{eqnarray}
where $\Gamma_0 \ll \epsilon$ is assumed. 

\section{\label{appendix_LS}Least squares fit}

Technical details of the least squares fits are summarized in this section. 
The computations of the least squares fits were performed 
on the CX400 supercomputer \cite{NISTdisclaimer} 
using a non-linear least squares program \cite{Kadowaki2018githubLS} 
based on the Levenberg-Marquardt algorithm. 
The difficulty of the present minimization problem of $\chi^2$ [Eq.~(\ref{chi2})] 
is caused by a fact that $\chi^2$ has many local minima in the parameter space. 
A trivial origin of this difficulty is that infinitesimal changes of 
$J_{\text{nn,eff}} \rightarrow (1+\epsilon) J_{\text{nn,eff}}$, 
where $J_{\text{nn,eff}}= J_1 + \tfrac{3}{5}D$ ($>0$) is the effective 
ferromagnetic NN coupling for small $J_m$ ($m \ge 2$) \cite{Hertog00}, 
$J_m \rightarrow (1+\epsilon) J_m$ ($m \ge 2$), 
and $\chi_{\text{L}} \rightarrow (1-\epsilon) \chi_{\text{L}}$ in Eq.~(\ref{Gsus}) 
bring about $[S(\bm{Q})]_{\text{el}} \rightarrow (1-\epsilon) [S(\bm{Q})]_{\text{el}}$ 
[Eq.~(\ref{SQ_QEA})], 
and consequently do not change the $\bm{Q}$ dependence of $[S(\bm{Q})]_{\text{el}}$. 
To avoid the (nearly) rank deficiency in the QR decomposition due to this fact, 
we fixed $J_1$ in performing the least squares fits. 
Indications of occurrence of this problem can be seen as several ranges 
of $\chi^2 \simeq \text{const.}$ in the curves of 
Figs.~\ref{chisq_QSL_IN5_0p1K0p7K}, \ref{chisq_QO_IN5_0p1K0p7K}, and \ref{chisq_QO_AMATERAS}. 
In addition, there were other unknown origins for the many local minima. 
These difficulties could be avoided by introducing a weak constraint 
of the parameters, i.e., adding the penalty function 
$\sum_{2 \leq m \leq m_{\text{max}}} \left( \tfrac{J_m}{1 \text{ K}} \right)^8$ 
to $\chi^2$. 
This penalty function weakly restricts $J_m$ in the range $|J_m|<1$ K, 
which is a reasonable assumption, 
and can be treated in the framework of the Levenberg-Marquardt algorithm. 
By inspecting results of the least squares fits, 
we can conclude that sufficiently accurate solutions of the minimization problem 
were obtained for the present purpose \cite{Supplemental_Material}. 
The uncertainty of the typical coupling constants listed in Table~\ref{FitJ1J14} 
is of the order 0.1 K \cite{Supplemental_Material}.

\bibliography{TTO_HK_p1}

\begin{thebibliography}{55}%
\makeatletter
\providecommand \@ifxundefined [1]{%
 \@ifx{#1\undefined}
}%
\providecommand \@ifnum [1]{%
 \ifnum #1\expandafter \@firstoftwo
 \else \expandafter \@secondoftwo
 \fi
}%
\providecommand \@ifx [1]{%
 \ifx #1\expandafter \@firstoftwo
 \else \expandafter \@secondoftwo
 \fi
}%
\providecommand \natexlab [1]{#1}%
\providecommand \enquote  [1]{``#1''}%
\providecommand \bibnamefont  [1]{#1}%
\providecommand \bibfnamefont [1]{#1}%
\providecommand \citenamefont [1]{#1}%
\providecommand \href@noop [0]{\@secondoftwo}%
\providecommand \href [0]{\begingroup \@sanitize@url \@href}%
\providecommand \@href[1]{\@@startlink{#1}\@@href}%
\providecommand \@@href[1]{\endgroup#1\@@endlink}%
\providecommand \@sanitize@url [0]{\catcode `\\12\catcode `\$12\catcode
  `\&12\catcode `\#12\catcode `\^12\catcode `\_12\catcode `\%12\relax}%
\providecommand \@@startlink[1]{}%
\providecommand \@@endlink[0]{}%
\providecommand \url  [0]{\begingroup\@sanitize@url \@url }%
\providecommand \@url [1]{\endgroup\@href {#1}{\urlprefix }}%
\providecommand \urlprefix  [0]{URL }%
\providecommand \Eprint [0]{\href }%
\providecommand \doibase [0]{http://dx.doi.org/}%
\providecommand \selectlanguage [0]{\@gobble}%
\providecommand \bibinfo  [0]{\@secondoftwo}%
\providecommand \bibfield  [0]{\@secondoftwo}%
\providecommand \translation [1]{[#1]}%
\providecommand \BibitemOpen [0]{}%
\providecommand \bibitemStop [0]{}%
\providecommand \bibitemNoStop [0]{.\EOS\space}%
\providecommand \EOS [0]{\spacefactor3000\relax}%
\providecommand \BibitemShut  [1]{\csname bibitem#1\endcsname}%
\let\auto@bib@innerbib\@empty
%</preamble>
\bibitem [{\citenamefont {Wannier}(1950)}]{Wannier50}%
  \BibitemOpen
  \bibfield  {author} {\bibinfo {author} {\bibfnamefont {G.~H.}\ \bibnamefont
  {Wannier}},\ }\href {\doibase 10.1103/PhysRev.79.357} {\bibfield  {journal}
  {\bibinfo  {journal} {Phys. Rev.}\ }\textbf {\bibinfo {volume} {79}},\
  \bibinfo {pages} {357} (\bibinfo {year} {1950})}\BibitemShut {NoStop}%
\bibitem [{\citenamefont {{Sy\^ozi}}(1951)}]{Shyozi51}%
  \BibitemOpen
  \bibfield  {author} {\bibinfo {author} {\bibfnamefont {I.}~\bibnamefont
  {{Sy\^ozi}}},\ }\href@noop {} {\bibfield  {journal} {\bibinfo  {journal}
  {Prog. Theor. Phys.}\ }\textbf {\bibinfo {volume} {6}},\ \bibinfo {pages}
  {306} (\bibinfo {year} {1951})}\BibitemShut {NoStop}%
\bibitem [{\citenamefont {Qi}\ \emph {et~al.}(2008)\citenamefont {Qi},
  \citenamefont {Brintlinger},\ and\ \citenamefont {Cumings}}]{Qi2008}%
  \BibitemOpen
  \bibfield  {author} {\bibinfo {author} {\bibfnamefont {Y.}~\bibnamefont
  {Qi}}, \bibinfo {author} {\bibfnamefont {T.}~\bibnamefont {Brintlinger}}, \
  and\ \bibinfo {author} {\bibfnamefont {J.}~\bibnamefont {Cumings}},\ }\href
  {\doibase 10.1103/PhysRevB.77.094418} {\bibfield  {journal} {\bibinfo
  {journal} {Phys. Rev. B}\ }\textbf {\bibinfo {volume} {77}},\ \bibinfo
  {pages} {094418} (\bibinfo {year} {2008})}\BibitemShut {NoStop}%
\bibitem [{\citenamefont {Gardner}\ \emph {et~al.}(2010)\citenamefont
  {Gardner}, \citenamefont {Gingras},\ and\ \citenamefont
  {Greedan}}]{Gardner10}%
  \BibitemOpen
  \bibfield  {author} {\bibinfo {author} {\bibfnamefont {J.~S.}\ \bibnamefont
  {Gardner}}, \bibinfo {author} {\bibfnamefont {M.~J.~P.}\ \bibnamefont
  {Gingras}}, \ and\ \bibinfo {author} {\bibfnamefont {J.~E.}\ \bibnamefont
  {Greedan}},\ }\href {\doibase 10.1103/RevModPhys.82.53} {\bibfield  {journal}
  {\bibinfo  {journal} {Rev. Mod. Phys.}\ }\textbf {\bibinfo {volume} {82}},\
  \bibinfo {pages} {53} (\bibinfo {year} {2010})}\BibitemShut {NoStop}%
\bibitem [{\citenamefont {Lacroix}\ \emph {et~al.}(2011)\citenamefont
  {Lacroix}, \citenamefont {Mendels},\ and\ \citenamefont {Mila}}]{Lacroix11}%
  \BibitemOpen
  \bibinfo {editor} {\bibfnamefont {C.}~\bibnamefont {Lacroix}}, \bibinfo
  {editor} {\bibfnamefont {P.}~\bibnamefont {Mendels}}, \ and\ \bibinfo
  {editor} {\bibfnamefont {F.}~\bibnamefont {Mila}},\ eds.,\ \href@noop {}
  {\emph {\bibinfo {title} {Introduction to Frustrated Magnetism}}}\ (\bibinfo
  {publisher} {Springer, Berlin, Heidelberg},\ \bibinfo {year}
  {2011})\BibitemShut {NoStop}%
\bibitem [{\citenamefont {{Bramwell}}\ and\ \citenamefont
  {{Gingras}}(2001)}]{Bramwell01}%
  \BibitemOpen
  \bibfield  {author} {\bibinfo {author} {\bibfnamefont {S.~T.}\ \bibnamefont
  {{Bramwell}}}\ and\ \bibinfo {author} {\bibfnamefont {M.~J.~P.}\ \bibnamefont
  {{Gingras}}},\ }\href@noop {} {\bibfield  {journal} {\bibinfo  {journal}
  {Science}\ }\textbf {\bibinfo {volume} {294}},\ \bibinfo {pages} {1495}
  (\bibinfo {year} {2001})}\BibitemShut {NoStop}%
\bibitem [{\citenamefont {{Ramirez}}\ \emph {et~al.}(1999)\citenamefont
  {{Ramirez}}, \citenamefont {{Hayashi}}, \citenamefont {{Cava}}, \citenamefont
  {{Siddharthan}},\ and\ \citenamefont {{Shastry}}}]{Ramirez99}%
  \BibitemOpen
  \bibfield  {author} {\bibinfo {author} {\bibfnamefont {A.~P.}\ \bibnamefont
  {{Ramirez}}}, \bibinfo {author} {\bibfnamefont {A.}~\bibnamefont
  {{Hayashi}}}, \bibinfo {author} {\bibfnamefont {R.~J.}\ \bibnamefont
  {{Cava}}}, \bibinfo {author} {\bibfnamefont {R.}~\bibnamefont
  {{Siddharthan}}}, \ and\ \bibinfo {author} {\bibfnamefont {B.~S.}\
  \bibnamefont {{Shastry}}},\ }\href {\doibase 10.1038/20619} {\bibfield
  {journal} {\bibinfo  {journal} {\nat}\ }\textbf {\bibinfo {volume} {399}},\
  \bibinfo {pages} {333} (\bibinfo {year} {1999})}\BibitemShut {NoStop}%
\bibitem [{\citenamefont {{Matsuhira}}\ \emph {et~al.}(2002)\citenamefont
  {{Matsuhira}}, \citenamefont {{Hiroi}}, \citenamefont {{Tayama}},
  \citenamefont {{Takagi}},\ and\ \citenamefont
  {{Sakakibara}}}]{MatsuhiraJPCM2002}%
  \BibitemOpen
  \bibfield  {author} {\bibinfo {author} {\bibfnamefont {K.}~\bibnamefont
  {{Matsuhira}}}, \bibinfo {author} {\bibfnamefont {Z.}~\bibnamefont
  {{Hiroi}}}, \bibinfo {author} {\bibfnamefont {T.}~\bibnamefont {{Tayama}}},
  \bibinfo {author} {\bibfnamefont {S.}~\bibnamefont {{Takagi}}}, \ and\
  \bibinfo {author} {\bibfnamefont {T.}~\bibnamefont {{Sakakibara}}},\ }\href
  {\doibase 10.1088/0953-8984/14/29/101} {\bibfield  {journal} {\bibinfo
  {journal} {J. Phys. Condens. Matter}\ }\textbf {\bibinfo {volume} {14}},\
  \bibinfo {pages} {L559} (\bibinfo {year} {2002})}\BibitemShut {NoStop}%
\bibitem [{\citenamefont {{Castelnovo}}\ \emph {et~al.}(2008)\citenamefont
  {{Castelnovo}}, \citenamefont {{Moessner}},\ and\ \citenamefont
  {{Sondhi}}}]{Castelnovo08}%
  \BibitemOpen
  \bibfield  {author} {\bibinfo {author} {\bibfnamefont {C.}~\bibnamefont
  {{Castelnovo}}}, \bibinfo {author} {\bibfnamefont {R.}~\bibnamefont
  {{Moessner}}}, \ and\ \bibinfo {author} {\bibfnamefont {S.~L.}\ \bibnamefont
  {{Sondhi}}},\ }\href {\doibase 10.1038/nature06433} {\bibfield  {journal}
  {\bibinfo  {journal} {Nature}\ }\textbf {\bibinfo {volume} {451}},\ \bibinfo
  {pages} {42} (\bibinfo {year} {2008})}\BibitemShut {NoStop}%
\bibitem [{\citenamefont {Kadowaki}\ \emph {et~al.}(2009)\citenamefont
  {Kadowaki}, \citenamefont {Doi}, \citenamefont {Aoki}, \citenamefont
  {Tabata}, \citenamefont {Sato}, \citenamefont {Lynn}, \citenamefont
  {Matsuhira},\ and\ \citenamefont {Hiroi}}]{Kadowaki09}%
  \BibitemOpen
  \bibfield  {author} {\bibinfo {author} {\bibfnamefont {H.}~\bibnamefont
  {Kadowaki}}, \bibinfo {author} {\bibfnamefont {N.}~\bibnamefont {Doi}},
  \bibinfo {author} {\bibfnamefont {Y.}~\bibnamefont {Aoki}}, \bibinfo {author}
  {\bibfnamefont {Y.}~\bibnamefont {Tabata}}, \bibinfo {author} {\bibfnamefont
  {T.~J.}\ \bibnamefont {Sato}}, \bibinfo {author} {\bibfnamefont {J.~W.}\
  \bibnamefont {Lynn}}, \bibinfo {author} {\bibfnamefont {K.}~\bibnamefont
  {Matsuhira}}, \ and\ \bibinfo {author} {\bibfnamefont {Z.}~\bibnamefont
  {Hiroi}},\ }\href {\doibase 10.1143/JPSJ.78.103706} {\bibfield  {journal}
  {\bibinfo  {journal} {J. Phys. Soc. Jpn.}\ }\textbf {\bibinfo {volume}
  {78}},\ \bibinfo {pages} {103706} (\bibinfo {year} {2009})}\BibitemShut
  {NoStop}%
\bibitem [{\citenamefont {Kadowaki}\ and\ \citenamefont
  {Nishimori}(1998)}]{KadowakiNishimori1998}%
  \BibitemOpen
  \bibfield  {author} {\bibinfo {author} {\bibfnamefont {T.}~\bibnamefont
  {Kadowaki}}\ and\ \bibinfo {author} {\bibfnamefont {H.}~\bibnamefont
  {Nishimori}},\ }\href {\doibase 10.1103/PhysRevE.58.5355} {\bibfield
  {journal} {\bibinfo  {journal} {Phys. Rev. E}\ }\textbf {\bibinfo {volume}
  {58}},\ \bibinfo {pages} {5355} (\bibinfo {year} {1998})}\BibitemShut
  {NoStop}%
\bibitem [{\citenamefont {{King}}\ \emph {et~al.}(2018)\citenamefont {{King}},
  \citenamefont {{Carrasquilla}}, \citenamefont {{Ozfidan}}, \citenamefont
  {{Raymond}}, \citenamefont {{Andriyash}}, \citenamefont {{Berkley}},
  \citenamefont {{Reis}}, \citenamefont {{Lanting}}, \citenamefont {{Harris}},
  \citenamefont {{Poulin-Lamarre}}, \citenamefont {{Smirnov}}, \citenamefont
  {{Rich}}, \citenamefont {{Altomare}}, \citenamefont {{Bunyk}}, \citenamefont
  {{Whittaker}}, \citenamefont {{Swenson}}, \citenamefont {{Hoskinson}},
  \citenamefont {{Sato}}, \citenamefont {{Volkmann}}, \citenamefont
  {{Ladizinsky}}, \citenamefont {{Johnson}}, \citenamefont {{Hilton}},\ and\
  \citenamefont {{Amin}}}]{King2018}%
  \BibitemOpen
  \bibfield  {author} {\bibinfo {author} {\bibfnamefont {A.~D.}\ \bibnamefont
  {{King}}}, \bibinfo {author} {\bibfnamefont {J.}~\bibnamefont
  {{Carrasquilla}}}, \bibinfo {author} {\bibfnamefont {I.}~\bibnamefont
  {{Ozfidan}}}, \bibinfo {author} {\bibfnamefont {J.}~\bibnamefont
  {{Raymond}}}, \bibinfo {author} {\bibfnamefont {E.}~\bibnamefont
  {{Andriyash}}}, \bibinfo {author} {\bibfnamefont {A.}~\bibnamefont
  {{Berkley}}}, \bibinfo {author} {\bibfnamefont {M.}~\bibnamefont {{Reis}}},
  \bibinfo {author} {\bibfnamefont {T.~M.}\ \bibnamefont {{Lanting}}}, \bibinfo
  {author} {\bibfnamefont {R.}~\bibnamefont {{Harris}}}, \bibinfo {author}
  {\bibfnamefont {G.}~\bibnamefont {{Poulin-Lamarre}}}, \bibinfo {author}
  {\bibfnamefont {A.~Y.}\ \bibnamefont {{Smirnov}}}, \bibinfo {author}
  {\bibfnamefont {C.}~\bibnamefont {{Rich}}}, \bibinfo {author} {\bibfnamefont
  {F.}~\bibnamefont {{Altomare}}}, \bibinfo {author} {\bibfnamefont
  {P.}~\bibnamefont {{Bunyk}}}, \bibinfo {author} {\bibfnamefont
  {J.}~\bibnamefont {{Whittaker}}}, \bibinfo {author} {\bibfnamefont
  {L.}~\bibnamefont {{Swenson}}}, \bibinfo {author} {\bibfnamefont
  {E.}~\bibnamefont {{Hoskinson}}}, \bibinfo {author} {\bibfnamefont
  {Y.}~\bibnamefont {{Sato}}}, \bibinfo {author} {\bibfnamefont
  {M.}~\bibnamefont {{Volkmann}}}, \bibinfo {author} {\bibfnamefont
  {E.}~\bibnamefont {{Ladizinsky}}}, \bibinfo {author} {\bibfnamefont
  {M.}~\bibnamefont {{Johnson}}}, \bibinfo {author} {\bibfnamefont
  {J.}~\bibnamefont {{Hilton}}}, \ and\ \bibinfo {author} {\bibfnamefont
  {M.~H.}\ \bibnamefont {{Amin}}},\ }\href {\doibase 10.1038/s41586-018-0410-x}
  {\bibfield  {journal} {\bibinfo  {journal} {\nat}\ }\textbf {\bibinfo
  {volume} {560}},\ \bibinfo {pages} {456} (\bibinfo {year}
  {2018})}\BibitemShut {NoStop}%
\bibitem [{\citenamefont {Savary}\ and\ \citenamefont
  {Balents}(2017)}]{Savary2017}%
  \BibitemOpen
  \bibfield  {author} {\bibinfo {author} {\bibfnamefont {L.}~\bibnamefont
  {Savary}}\ and\ \bibinfo {author} {\bibfnamefont {L.}~\bibnamefont
  {Balents}},\ }\href@noop {} {\bibfield  {journal} {\bibinfo  {journal} {Rep.
  Prog. Phys.}\ }\textbf {\bibinfo {volume} {80}},\ \bibinfo {pages} {016502}
  (\bibinfo {year} {2017})}\BibitemShut {NoStop}%
\bibitem [{\citenamefont {Anderson}(1973)}]{Anderson73}%
  \BibitemOpen
  \bibfield  {author} {\bibinfo {author} {\bibfnamefont {P.~W.}\ \bibnamefont
  {Anderson}},\ }\href@noop {} {\bibfield  {journal} {\bibinfo  {journal}
  {Mater. Res. Bull.}\ }\textbf {\bibinfo {volume} {8}},\ \bibinfo {pages}
  {153} (\bibinfo {year} {1973})}\BibitemShut {NoStop}%
\bibitem [{\citenamefont {Hirakawa}\ \emph {et~al.}(1985)\citenamefont
  {Hirakawa}, \citenamefont {Kadowaki},\ and\ \citenamefont
  {Ubukoshi}}]{Hirakawa1985}%
  \BibitemOpen
  \bibfield  {author} {\bibinfo {author} {\bibfnamefont {K.}~\bibnamefont
  {Hirakawa}}, \bibinfo {author} {\bibfnamefont {H.}~\bibnamefont {Kadowaki}},
  \ and\ \bibinfo {author} {\bibfnamefont {K.}~\bibnamefont {Ubukoshi}},\
  }\href {\doibase 10.1143/JPSJ.54.3526} {\bibfield  {journal} {\bibinfo
  {journal} {J. Phys. Soc. Jpn.}\ }\textbf {\bibinfo {volume} {54}},\ \bibinfo
  {pages} {3526} (\bibinfo {year} {1985})}\BibitemShut {NoStop}%
\bibitem [{\citenamefont {Gardner}\ \emph {et~al.}(1999)\citenamefont
  {Gardner}, \citenamefont {Dunsiger}, \citenamefont {Gaulin}, \citenamefont
  {Gingras}, \citenamefont {Greedan}, \citenamefont {Kiefl}, \citenamefont
  {Lumsden}, \citenamefont {MacFarlane}, \citenamefont {Raju}, \citenamefont
  {Sonier}, \citenamefont {Swainson},\ and\ \citenamefont {Tun}}]{Gardner99}%
  \BibitemOpen
  \bibfield  {author} {\bibinfo {author} {\bibfnamefont {J.~S.}\ \bibnamefont
  {Gardner}}, \bibinfo {author} {\bibfnamefont {S.~R.}\ \bibnamefont
  {Dunsiger}}, \bibinfo {author} {\bibfnamefont {B.~D.}\ \bibnamefont
  {Gaulin}}, \bibinfo {author} {\bibfnamefont {M.~J.~P.}\ \bibnamefont
  {Gingras}}, \bibinfo {author} {\bibfnamefont {J.~E.}\ \bibnamefont
  {Greedan}}, \bibinfo {author} {\bibfnamefont {R.~F.}\ \bibnamefont {Kiefl}},
  \bibinfo {author} {\bibfnamefont {M.~D.}\ \bibnamefont {Lumsden}}, \bibinfo
  {author} {\bibfnamefont {W.~A.}\ \bibnamefont {MacFarlane}}, \bibinfo
  {author} {\bibfnamefont {N.~P.}\ \bibnamefont {Raju}}, \bibinfo {author}
  {\bibfnamefont {J.~E.}\ \bibnamefont {Sonier}}, \bibinfo {author}
  {\bibfnamefont {I.}~\bibnamefont {Swainson}}, \ and\ \bibinfo {author}
  {\bibfnamefont {Z.}~\bibnamefont {Tun}},\ }\href {\doibase
  10.1103/PhysRevLett.82.1012} {\bibfield  {journal} {\bibinfo  {journal}
  {Phys. Rev. Lett.}\ }\textbf {\bibinfo {volume} {82}},\ \bibinfo {pages}
  {1012} (\bibinfo {year} {1999})}\BibitemShut {NoStop}%
\bibitem [{\citenamefont {{Han}}\ \emph {et~al.}(2012)\citenamefont {{Han}},
  \citenamefont {{Helton}}, \citenamefont {{Chu}}, \citenamefont {{Nocera}},
  \citenamefont {{Rodriguez-Rivera}}, \citenamefont {{Broholm}},\ and\
  \citenamefont {{Lee}}}]{Han2012}%
  \BibitemOpen
  \bibfield  {author} {\bibinfo {author} {\bibfnamefont {T.-H.}\ \bibnamefont
  {{Han}}}, \bibinfo {author} {\bibfnamefont {J.~S.}\ \bibnamefont {{Helton}}},
  \bibinfo {author} {\bibfnamefont {S.}~\bibnamefont {{Chu}}}, \bibinfo
  {author} {\bibfnamefont {D.~G.}\ \bibnamefont {{Nocera}}}, \bibinfo {author}
  {\bibfnamefont {J.~A.}\ \bibnamefont {{Rodriguez-Rivera}}}, \bibinfo {author}
  {\bibfnamefont {C.}~\bibnamefont {{Broholm}}}, \ and\ \bibinfo {author}
  {\bibfnamefont {Y.~S.}\ \bibnamefont {{Lee}}},\ }\href {\doibase
  10.1038/nature11659} {\bibfield  {journal} {\bibinfo  {journal} {\nat}\
  }\textbf {\bibinfo {volume} {492}},\ \bibinfo {pages} {406} (\bibinfo {year}
  {2012})}\BibitemShut {NoStop}%
\bibitem [{\citenamefont {Ross}\ \emph {et~al.}(2011)\citenamefont {Ross},
  \citenamefont {Savary}, \citenamefont {Gaulin},\ and\ \citenamefont
  {Balents}}]{Ross11}%
  \BibitemOpen
  \bibfield  {author} {\bibinfo {author} {\bibfnamefont {K.~A.}\ \bibnamefont
  {Ross}}, \bibinfo {author} {\bibfnamefont {L.}~\bibnamefont {Savary}},
  \bibinfo {author} {\bibfnamefont {B.~D.}\ \bibnamefont {Gaulin}}, \ and\
  \bibinfo {author} {\bibfnamefont {L.}~\bibnamefont {Balents}},\ }\href
  {\doibase 10.1103/PhysRevX.1.021002} {\bibfield  {journal} {\bibinfo
  {journal} {Phys. Rev. X}\ }\textbf {\bibinfo {volume} {1}},\ \bibinfo {pages}
  {021002} (\bibinfo {year} {2011})}\BibitemShut {NoStop}%
\bibitem [{\citenamefont {{Chang}}\ \emph {et~al.}(2012)\citenamefont
  {{Chang}}, \citenamefont {{Onoda}}, \citenamefont {{Su}}, \citenamefont
  {{Kao}}, \citenamefont {{Tsuei}}, \citenamefont {{Yasui}}, \citenamefont
  {{Kakurai}},\ and\ \citenamefont {{Lees}}}]{Chang12}%
  \BibitemOpen
  \bibfield  {author} {\bibinfo {author} {\bibfnamefont {L.-J.}\ \bibnamefont
  {{Chang}}}, \bibinfo {author} {\bibfnamefont {S.}~\bibnamefont {{Onoda}}},
  \bibinfo {author} {\bibfnamefont {Y.}~\bibnamefont {{Su}}}, \bibinfo {author}
  {\bibfnamefont {Y.-J.}\ \bibnamefont {{Kao}}}, \bibinfo {author}
  {\bibfnamefont {K.-D.}\ \bibnamefont {{Tsuei}}}, \bibinfo {author}
  {\bibfnamefont {Y.}~\bibnamefont {{Yasui}}}, \bibinfo {author} {\bibfnamefont
  {K.}~\bibnamefont {{Kakurai}}}, \ and\ \bibinfo {author} {\bibfnamefont
  {M.~R.}\ \bibnamefont {{Lees}}},\ }\href {\doibase 10.1038/ncomms1989}
  {\bibfield  {journal} {\bibinfo  {journal} {Nature Communications}\ }\textbf
  {\bibinfo {volume} {3}},\ \bibinfo {eid} {992} (\bibinfo {year}
  {2012})}\BibitemShut {NoStop}%
\bibitem [{\citenamefont {{Shen}}\ \emph {et~al.}(2016)\citenamefont {{Shen}},
  \citenamefont {{Li}}, \citenamefont {{Wo}}, \citenamefont {{Li}},
  \citenamefont {{Shen}}, \citenamefont {{Pan}}, \citenamefont {{Wang}},
  \citenamefont {{Walker}}, \citenamefont {{Steffens}}, \citenamefont
  {{Boehm}}, \citenamefont {{Hao}}, \citenamefont {{Quintero-Castro}},
  \citenamefont {{Harriger}}, \citenamefont {{Frontzek}}, \citenamefont
  {{Hao}}, \citenamefont {{Meng}}, \citenamefont {{Zhang}}, \citenamefont
  {{Chen}},\ and\ \citenamefont {{Zhao}}}]{Shen2016}%
  \BibitemOpen
  \bibfield  {author} {\bibinfo {author} {\bibfnamefont {Y.}~\bibnamefont
  {{Shen}}}, \bibinfo {author} {\bibfnamefont {Y.-D.}\ \bibnamefont {{Li}}},
  \bibinfo {author} {\bibfnamefont {H.}~\bibnamefont {{Wo}}}, \bibinfo {author}
  {\bibfnamefont {Y.}~\bibnamefont {{Li}}}, \bibinfo {author} {\bibfnamefont
  {S.}~\bibnamefont {{Shen}}}, \bibinfo {author} {\bibfnamefont
  {B.}~\bibnamefont {{Pan}}}, \bibinfo {author} {\bibfnamefont
  {Q.}~\bibnamefont {{Wang}}}, \bibinfo {author} {\bibfnamefont {H.~C.}\
  \bibnamefont {{Walker}}}, \bibinfo {author} {\bibfnamefont {P.}~\bibnamefont
  {{Steffens}}}, \bibinfo {author} {\bibfnamefont {M.}~\bibnamefont {{Boehm}}},
  \bibinfo {author} {\bibfnamefont {Y.}~\bibnamefont {{Hao}}}, \bibinfo
  {author} {\bibfnamefont {D.~L.}\ \bibnamefont {{Quintero-Castro}}}, \bibinfo
  {author} {\bibfnamefont {L.~W.}\ \bibnamefont {{Harriger}}}, \bibinfo
  {author} {\bibfnamefont {M.~D.}\ \bibnamefont {{Frontzek}}}, \bibinfo
  {author} {\bibfnamefont {L.}~\bibnamefont {{Hao}}}, \bibinfo {author}
  {\bibfnamefont {S.}~\bibnamefont {{Meng}}}, \bibinfo {author} {\bibfnamefont
  {Q.}~\bibnamefont {{Zhang}}}, \bibinfo {author} {\bibfnamefont
  {G.}~\bibnamefont {{Chen}}}, \ and\ \bibinfo {author} {\bibfnamefont
  {J.}~\bibnamefont {{Zhao}}},\ }\href {\doibase 10.1038/nature20614}
  {\bibfield  {journal} {\bibinfo  {journal} {\nat}\ }\textbf {\bibinfo
  {volume} {540}},\ \bibinfo {pages} {559} (\bibinfo {year}
  {2016})}\BibitemShut {NoStop}%
\bibitem [{\citenamefont {F\aa{}k}\ \emph {et~al.}(2017)\citenamefont
  {F\aa{}k}, \citenamefont {Bieri}, \citenamefont {Can\'evet}, \citenamefont
  {Messio}, \citenamefont {Payen}, \citenamefont {Viaud}, \citenamefont
  {Guillot-Deudon}, \citenamefont {Darie}, \citenamefont {Ollivier},\ and\
  \citenamefont {Mendels}}]{Fak2017}%
  \BibitemOpen
  \bibfield  {author} {\bibinfo {author} {\bibfnamefont {B.}~\bibnamefont
  {F\aa{}k}}, \bibinfo {author} {\bibfnamefont {S.}~\bibnamefont {Bieri}},
  \bibinfo {author} {\bibfnamefont {E.}~\bibnamefont {Can\'evet}}, \bibinfo
  {author} {\bibfnamefont {L.}~\bibnamefont {Messio}}, \bibinfo {author}
  {\bibfnamefont {C.}~\bibnamefont {Payen}}, \bibinfo {author} {\bibfnamefont
  {M.}~\bibnamefont {Viaud}}, \bibinfo {author} {\bibfnamefont
  {C.}~\bibnamefont {Guillot-Deudon}}, \bibinfo {author} {\bibfnamefont
  {C.}~\bibnamefont {Darie}}, \bibinfo {author} {\bibfnamefont
  {J.}~\bibnamefont {Ollivier}}, \ and\ \bibinfo {author} {\bibfnamefont
  {P.}~\bibnamefont {Mendels}},\ }\href {\doibase 10.1103/PhysRevB.95.060402}
  {\bibfield  {journal} {\bibinfo  {journal} {Phys. Rev. B}\ }\textbf {\bibinfo
  {volume} {95}},\ \bibinfo {pages} {060402} (\bibinfo {year}
  {2017})}\BibitemShut {NoStop}%
\bibitem [{\citenamefont {{Sibille}}\ \emph {et~al.}(2018)\citenamefont
  {{Sibille}}, \citenamefont {{Gauthier}}, \citenamefont {{Yan}}, \citenamefont
  {{Ciomaga Hatnean}}, \citenamefont {{Ollivier}}, \citenamefont {{Winn}},
  \citenamefont {{Filges}}, \citenamefont {{Balakrishnan}}, \citenamefont
  {{Kenzelmann}}, \citenamefont {{Shannon}},\ and\ \citenamefont
  {{Fennell}}}]{Sibille2018}%
  \BibitemOpen
  \bibfield  {author} {\bibinfo {author} {\bibfnamefont {R.}~\bibnamefont
  {{Sibille}}}, \bibinfo {author} {\bibfnamefont {N.}~\bibnamefont
  {{Gauthier}}}, \bibinfo {author} {\bibfnamefont {H.}~\bibnamefont {{Yan}}},
  \bibinfo {author} {\bibfnamefont {M.}~\bibnamefont {{Ciomaga Hatnean}}},
  \bibinfo {author} {\bibfnamefont {J.}~\bibnamefont {{Ollivier}}}, \bibinfo
  {author} {\bibfnamefont {B.}~\bibnamefont {{Winn}}}, \bibinfo {author}
  {\bibfnamefont {U.}~\bibnamefont {{Filges}}}, \bibinfo {author}
  {\bibfnamefont {G.}~\bibnamefont {{Balakrishnan}}}, \bibinfo {author}
  {\bibfnamefont {M.}~\bibnamefont {{Kenzelmann}}}, \bibinfo {author}
  {\bibfnamefont {N.}~\bibnamefont {{Shannon}}}, \ and\ \bibinfo {author}
  {\bibfnamefont {T.}~\bibnamefont {{Fennell}}},\ }\href {\doibase
  10.1038/s41567-018-0116-x} {\bibfield  {journal} {\bibinfo  {journal} {Nature
  Physics}\ }\textbf {\bibinfo {volume} {14}},\ \bibinfo {pages} {711}
  (\bibinfo {year} {2018})}\BibitemShut {NoStop}%
\bibitem [{\citenamefont {{Kadowaki}}\ \emph {et~al.}(2018)\citenamefont
  {{Kadowaki}}, \citenamefont {{Wakita}}, \citenamefont {{F{\aa}k}},
  \citenamefont {{Ollivier}}, \citenamefont {{Ohira-Kawamura}}, \citenamefont
  {{Nakajima}}, \citenamefont {{Takatsu}},\ and\ \citenamefont
  {{Tamai}}}]{Kadowaki2018}%
  \BibitemOpen
  \bibfield  {author} {\bibinfo {author} {\bibfnamefont {H.}~\bibnamefont
  {{Kadowaki}}}, \bibinfo {author} {\bibfnamefont {M.}~\bibnamefont
  {{Wakita}}}, \bibinfo {author} {\bibfnamefont {B.}~\bibnamefont {{F{\aa}k}}},
  \bibinfo {author} {\bibfnamefont {J.}~\bibnamefont {{Ollivier}}}, \bibinfo
  {author} {\bibfnamefont {S.}~\bibnamefont {{Ohira-Kawamura}}}, \bibinfo
  {author} {\bibfnamefont {K.}~\bibnamefont {{Nakajima}}}, \bibinfo {author}
  {\bibfnamefont {H.}~\bibnamefont {{Takatsu}}}, \ and\ \bibinfo {author}
  {\bibfnamefont {M.}~\bibnamefont {{Tamai}}},\ }\href {\doibase
  10.7566/JPSJ.87.064704} {\bibfield  {journal} {\bibinfo  {journal} {J. Phys.
  Soc. Jpn.}\ }\textbf {\bibinfo {volume} {87}},\ \bibinfo {eid} {064704}
  (\bibinfo {year} {2018})}\BibitemShut {NoStop}%
\bibitem [{\citenamefont {Molavian}\ \emph {et~al.}(2007)\citenamefont
  {Molavian}, \citenamefont {Gingras},\ and\ \citenamefont
  {Canals}}]{Molavian07}%
  \BibitemOpen
  \bibfield  {author} {\bibinfo {author} {\bibfnamefont {H.~R.}\ \bibnamefont
  {Molavian}}, \bibinfo {author} {\bibfnamefont {M.~J.~P.}\ \bibnamefont
  {Gingras}}, \ and\ \bibinfo {author} {\bibfnamefont {B.}~\bibnamefont
  {Canals}},\ }\href {\doibase 10.1103/PhysRevLett.98.157204} {\bibfield
  {journal} {\bibinfo  {journal} {Phys. Rev. Lett.}\ }\textbf {\bibinfo
  {volume} {98}},\ \bibinfo {pages} {157204} (\bibinfo {year}
  {2007})}\BibitemShut {NoStop}%
\bibitem [{\citenamefont {Gingras}\ and\ \citenamefont
  {McClarty}(2014)}]{Gingras14}%
  \BibitemOpen
  \bibfield  {author} {\bibinfo {author} {\bibfnamefont {M.~J.~P.}\
  \bibnamefont {Gingras}}\ and\ \bibinfo {author} {\bibfnamefont {P.~A.}\
  \bibnamefont {McClarty}},\ }\href
  {http://stacks.iop.org/0034-4885/77/i=5/a=056501} {\bibfield  {journal}
  {\bibinfo  {journal} {Rep. Prog. Phys.}\ }\textbf {\bibinfo {volume} {77}},\
  \bibinfo {pages} {056501} (\bibinfo {year} {2014})}\BibitemShut {NoStop}%
\bibitem [{\citenamefont {Taniguchi}\ \emph {et~al.}(2013)\citenamefont
  {Taniguchi}, \citenamefont {Kadowaki}, \citenamefont {Takatsu}, \citenamefont
  {F\aa{}k}, \citenamefont {Ollivier}, \citenamefont {Yamazaki}, \citenamefont
  {Sato}, \citenamefont {Yoshizawa}, \citenamefont {Shimura}, \citenamefont
  {Sakakibara}, \citenamefont {Hong}, \citenamefont {Goto}, \citenamefont
  {Yaraskavitch},\ and\ \citenamefont {Kycia}}]{Taniguchi13}%
  \BibitemOpen
  \bibfield  {author} {\bibinfo {author} {\bibfnamefont {T.}~\bibnamefont
  {Taniguchi}}, \bibinfo {author} {\bibfnamefont {H.}~\bibnamefont {Kadowaki}},
  \bibinfo {author} {\bibfnamefont {H.}~\bibnamefont {Takatsu}}, \bibinfo
  {author} {\bibfnamefont {B.}~\bibnamefont {F\aa{}k}}, \bibinfo {author}
  {\bibfnamefont {J.}~\bibnamefont {Ollivier}}, \bibinfo {author}
  {\bibfnamefont {T.}~\bibnamefont {Yamazaki}}, \bibinfo {author}
  {\bibfnamefont {T.~J.}\ \bibnamefont {Sato}}, \bibinfo {author}
  {\bibfnamefont {H.}~\bibnamefont {Yoshizawa}}, \bibinfo {author}
  {\bibfnamefont {Y.}~\bibnamefont {Shimura}}, \bibinfo {author} {\bibfnamefont
  {T.}~\bibnamefont {Sakakibara}}, \bibinfo {author} {\bibfnamefont
  {T.}~\bibnamefont {Hong}}, \bibinfo {author} {\bibfnamefont {K.}~\bibnamefont
  {Goto}}, \bibinfo {author} {\bibfnamefont {L.~R.}\ \bibnamefont
  {Yaraskavitch}}, \ and\ \bibinfo {author} {\bibfnamefont {J.~B.}\
  \bibnamefont {Kycia}},\ }\href {\doibase 10.1103/PhysRevB.87.060408}
  {\bibfield  {journal} {\bibinfo  {journal} {Phys. Rev. B}\ }\textbf {\bibinfo
  {volume} {87}},\ \bibinfo {pages} {060408} (\bibinfo {year}
  {2013})}\BibitemShut {NoStop}%
\bibitem [{\citenamefont {Wakita}\ \emph {et~al.}(2016)\citenamefont {Wakita},
  \citenamefont {Taniguchi}, \citenamefont {Edamoto}, \citenamefont {Takatsu},\
  and\ \citenamefont {Kadowaki}}]{Wakita2016}%
  \BibitemOpen
  \bibfield  {author} {\bibinfo {author} {\bibfnamefont {M.}~\bibnamefont
  {Wakita}}, \bibinfo {author} {\bibfnamefont {T.}~\bibnamefont {Taniguchi}},
  \bibinfo {author} {\bibfnamefont {H.}~\bibnamefont {Edamoto}}, \bibinfo
  {author} {\bibfnamefont {H.}~\bibnamefont {Takatsu}}, \ and\ \bibinfo
  {author} {\bibfnamefont {H.}~\bibnamefont {Kadowaki}},\ }\href
  {http://stacks.iop.org/1742-6596/683/i=1/a=012023} {\bibfield  {journal}
  {\bibinfo  {journal} {J. Phys.: Conf. Series}\ }\textbf {\bibinfo {volume}
  {683}},\ \bibinfo {pages} {012023} (\bibinfo {year} {2016})}\BibitemShut
  {NoStop}%
\bibitem [{\citenamefont {Takatsu}\ \emph
  {et~al.}(2016{\natexlab{a}})\citenamefont {Takatsu}, \citenamefont {Onoda},
  \citenamefont {Kittaka}, \citenamefont {Kasahara}, \citenamefont {Kono},
  \citenamefont {Sakakibara}, \citenamefont {Kato}, \citenamefont {F\aa{}k},
  \citenamefont {Ollivier}, \citenamefont {Lynn}, \citenamefont {Taniguchi},
  \citenamefont {Wakita},\ and\ \citenamefont {Kadowaki}}]{Takatsu2016prl}%
  \BibitemOpen
  \bibfield  {author} {\bibinfo {author} {\bibfnamefont {H.}~\bibnamefont
  {Takatsu}}, \bibinfo {author} {\bibfnamefont {S.}~\bibnamefont {Onoda}},
  \bibinfo {author} {\bibfnamefont {S.}~\bibnamefont {Kittaka}}, \bibinfo
  {author} {\bibfnamefont {A.}~\bibnamefont {Kasahara}}, \bibinfo {author}
  {\bibfnamefont {Y.}~\bibnamefont {Kono}}, \bibinfo {author} {\bibfnamefont
  {T.}~\bibnamefont {Sakakibara}}, \bibinfo {author} {\bibfnamefont
  {Y.}~\bibnamefont {Kato}}, \bibinfo {author} {\bibfnamefont {B.}~\bibnamefont
  {F\aa{}k}}, \bibinfo {author} {\bibfnamefont {J.}~\bibnamefont {Ollivier}},
  \bibinfo {author} {\bibfnamefont {J.~W.}\ \bibnamefont {Lynn}}, \bibinfo
  {author} {\bibfnamefont {T.}~\bibnamefont {Taniguchi}}, \bibinfo {author}
  {\bibfnamefont {M.}~\bibnamefont {Wakita}}, \ and\ \bibinfo {author}
  {\bibfnamefont {H.}~\bibnamefont {Kadowaki}},\ }\href {\doibase
  10.1103/PhysRevLett.116.217201} {\bibfield  {journal} {\bibinfo  {journal}
  {Phys. Rev. Lett.}\ }\textbf {\bibinfo {volume} {116}},\ \bibinfo {pages}
  {217201} (\bibinfo {year} {2016}{\natexlab{a}})}\BibitemShut {NoStop}%
\bibitem [{\citenamefont {Takatsu}\ \emph
  {et~al.}(2016{\natexlab{b}})\citenamefont {Takatsu}, \citenamefont
  {Taniguchi}, \citenamefont {Kittaka}, \citenamefont {Sakakibara},\ and\
  \citenamefont {Kadowaki}}]{Takatsu2016JPCS}%
  \BibitemOpen
  \bibfield  {author} {\bibinfo {author} {\bibfnamefont {H.}~\bibnamefont
  {Takatsu}}, \bibinfo {author} {\bibfnamefont {T.}~\bibnamefont {Taniguchi}},
  \bibinfo {author} {\bibfnamefont {S.}~\bibnamefont {Kittaka}}, \bibinfo
  {author} {\bibfnamefont {T.}~\bibnamefont {Sakakibara}}, \ and\ \bibinfo
  {author} {\bibfnamefont {H.}~\bibnamefont {Kadowaki}},\ }\href
  {http://stacks.iop.org/1742-6596/683/i=1/a=012022} {\bibfield  {journal}
  {\bibinfo  {journal} {J. Phys.: Conf. Series}\ }\textbf {\bibinfo {volume}
  {683}},\ \bibinfo {pages} {012022} (\bibinfo {year}
  {2016}{\natexlab{b}})}\BibitemShut {NoStop}%
\bibitem [{\citenamefont {Kadowaki}\ \emph {et~al.}(2018)\citenamefont
  {Kadowaki}, \citenamefont {Takatsu},\ and\ \citenamefont
  {Wakita}}]{Kadowaki2018prb}%
  \BibitemOpen
  \bibfield  {author} {\bibinfo {author} {\bibfnamefont {H.}~\bibnamefont
  {Kadowaki}}, \bibinfo {author} {\bibfnamefont {H.}~\bibnamefont {Takatsu}}, \
  and\ \bibinfo {author} {\bibfnamefont {M.}~\bibnamefont {Wakita}},\ }\href
  {\doibase 10.1103/PhysRevB.98.144410} {\bibfield  {journal} {\bibinfo
  {journal} {Phys. Rev. B}\ }\textbf {\bibinfo {volume} {98}},\ \bibinfo
  {pages} {144410} (\bibinfo {year} {2018})}\BibitemShut {NoStop}%
\bibitem [{\citenamefont {Onoda}\ and\ \citenamefont {Tanaka}(2011)}]{Onoda11}%
  \BibitemOpen
  \bibfield  {author} {\bibinfo {author} {\bibfnamefont {S.}~\bibnamefont
  {Onoda}}\ and\ \bibinfo {author} {\bibfnamefont {Y.}~\bibnamefont {Tanaka}},\
  }\href {\doibase 10.1103/PhysRevB.83.094411} {\bibfield  {journal} {\bibinfo
  {journal} {Phys. Rev. B}\ }\textbf {\bibinfo {volume} {83}},\ \bibinfo
  {pages} {094411} (\bibinfo {year} {2011})}\BibitemShut {NoStop}%
\bibitem [{\citenamefont {Lee}\ \emph {et~al.}(2012)\citenamefont {Lee},
  \citenamefont {Onoda},\ and\ \citenamefont {Balents}}]{Lee12}%
  \BibitemOpen
  \bibfield  {author} {\bibinfo {author} {\bibfnamefont {S.}~\bibnamefont
  {Lee}}, \bibinfo {author} {\bibfnamefont {S.}~\bibnamefont {Onoda}}, \ and\
  \bibinfo {author} {\bibfnamefont {L.}~\bibnamefont {Balents}},\ }\href
  {\doibase 10.1103/PhysRevB.86.104412} {\bibfield  {journal} {\bibinfo
  {journal} {Phys. Rev. B}\ }\textbf {\bibinfo {volume} {86}},\ \bibinfo
  {pages} {104412} (\bibinfo {year} {2012})}\BibitemShut {NoStop}%
\bibitem [{\citenamefont {Hermele}\ \emph {et~al.}(2004)\citenamefont
  {Hermele}, \citenamefont {Fisher},\ and\ \citenamefont
  {Balents}}]{Hermele04}%
  \BibitemOpen
  \bibfield  {author} {\bibinfo {author} {\bibfnamefont {M.}~\bibnamefont
  {Hermele}}, \bibinfo {author} {\bibfnamefont {M.~P.~A.}\ \bibnamefont
  {Fisher}}, \ and\ \bibinfo {author} {\bibfnamefont {L.}~\bibnamefont
  {Balents}},\ }\href {\doibase 10.1103/PhysRevB.69.064404} {\bibfield
  {journal} {\bibinfo  {journal} {Phys. Rev. B}\ }\textbf {\bibinfo {volume}
  {69}},\ \bibinfo {pages} {064404} (\bibinfo {year} {2004})}\BibitemShut
  {NoStop}%
\bibitem [{\citenamefont {{Yasui}}\ \emph {et~al.}(2002)\citenamefont
  {{Yasui}}, \citenamefont {{Kanada}}, \citenamefont {{Ito}}, \citenamefont
  {{Harashina}}, \citenamefont {{Sato}}, \citenamefont {{Okumura}},
  \citenamefont {{Kakurai}},\ and\ \citenamefont {{Kadowaki}}}]{Yasui2002}%
  \BibitemOpen
  \bibfield  {author} {\bibinfo {author} {\bibfnamefont {Y.}~\bibnamefont
  {{Yasui}}}, \bibinfo {author} {\bibfnamefont {M.}~\bibnamefont {{Kanada}}},
  \bibinfo {author} {\bibfnamefont {M.}~\bibnamefont {{Ito}}}, \bibinfo
  {author} {\bibfnamefont {H.}~\bibnamefont {{Harashina}}}, \bibinfo {author}
  {\bibfnamefont {M.}~\bibnamefont {{Sato}}}, \bibinfo {author} {\bibfnamefont
  {H.}~\bibnamefont {{Okumura}}}, \bibinfo {author} {\bibfnamefont
  {K.}~\bibnamefont {{Kakurai}}}, \ and\ \bibinfo {author} {\bibfnamefont
  {H.}~\bibnamefont {{Kadowaki}}},\ }\href {\doibase 10.1143/JPSJ.71.599}
  {\bibfield  {journal} {\bibinfo  {journal} {J. Phys. Soc. Jpn.}\ }\textbf
  {\bibinfo {volume} {71}},\ \bibinfo {pages} {599} (\bibinfo {year}
  {2002})}\BibitemShut {NoStop}%
\bibitem [{\citenamefont {Fennell}\ \emph {et~al.}(2012)\citenamefont
  {Fennell}, \citenamefont {Kenzelmann}, \citenamefont {Roessli}, \citenamefont
  {Haas},\ and\ \citenamefont {Cava}}]{Fennell2012}%
  \BibitemOpen
  \bibfield  {author} {\bibinfo {author} {\bibfnamefont {T.}~\bibnamefont
  {Fennell}}, \bibinfo {author} {\bibfnamefont {M.}~\bibnamefont {Kenzelmann}},
  \bibinfo {author} {\bibfnamefont {B.}~\bibnamefont {Roessli}}, \bibinfo
  {author} {\bibfnamefont {M.~K.}\ \bibnamefont {Haas}}, \ and\ \bibinfo
  {author} {\bibfnamefont {R.~J.}\ \bibnamefont {Cava}},\ }\href {\doibase
  10.1103/PhysRevLett.109.017201} {\bibfield  {journal} {\bibinfo  {journal}
  {Phys. Rev. Lett.}\ }\textbf {\bibinfo {volume} {109}},\ \bibinfo {pages}
  {017201} (\bibinfo {year} {2012})}\BibitemShut {NoStop}%
\bibitem [{\citenamefont {Petit}\ \emph {et~al.}(2012)\citenamefont {Petit},
  \citenamefont {Bonville}, \citenamefont {Robert}, \citenamefont {Decorse},\
  and\ \citenamefont {Mirebeau}}]{Petit12}%
  \BibitemOpen
  \bibfield  {author} {\bibinfo {author} {\bibfnamefont {S.}~\bibnamefont
  {Petit}}, \bibinfo {author} {\bibfnamefont {P.}~\bibnamefont {Bonville}},
  \bibinfo {author} {\bibfnamefont {J.}~\bibnamefont {Robert}}, \bibinfo
  {author} {\bibfnamefont {C.}~\bibnamefont {Decorse}}, \ and\ \bibinfo
  {author} {\bibfnamefont {I.}~\bibnamefont {Mirebeau}},\ }\href {\doibase
  10.1103/PhysRevB.86.174403} {\bibfield  {journal} {\bibinfo  {journal} {Phys.
  Rev. B}\ }\textbf {\bibinfo {volume} {86}},\ \bibinfo {pages} {174403}
  (\bibinfo {year} {2012})}\BibitemShut {NoStop}%
\bibitem [{\citenamefont {Fritsch}\ \emph {et~al.}(2013)\citenamefont
  {Fritsch}, \citenamefont {Ross}, \citenamefont {Qiu}, \citenamefont {Copley},
  \citenamefont {Guidi}, \citenamefont {Bewley}, \citenamefont {Dabkowska},\
  and\ \citenamefont {Gaulin}}]{Fritsch13}%
  \BibitemOpen
  \bibfield  {author} {\bibinfo {author} {\bibfnamefont {K.}~\bibnamefont
  {Fritsch}}, \bibinfo {author} {\bibfnamefont {K.~A.}\ \bibnamefont {Ross}},
  \bibinfo {author} {\bibfnamefont {Y.}~\bibnamefont {Qiu}}, \bibinfo {author}
  {\bibfnamefont {J.~R.~D.}\ \bibnamefont {Copley}}, \bibinfo {author}
  {\bibfnamefont {T.}~\bibnamefont {Guidi}}, \bibinfo {author} {\bibfnamefont
  {R.~I.}\ \bibnamefont {Bewley}}, \bibinfo {author} {\bibfnamefont {H.~A.}\
  \bibnamefont {Dabkowska}}, \ and\ \bibinfo {author} {\bibfnamefont {B.~D.}\
  \bibnamefont {Gaulin}},\ }\href {\doibase 10.1103/PhysRevB.87.094410}
  {\bibfield  {journal} {\bibinfo  {journal} {Phys. Rev. B}\ }\textbf {\bibinfo
  {volume} {87}},\ \bibinfo {pages} {094410} (\bibinfo {year}
  {2013})}\BibitemShut {NoStop}%
\bibitem [{\citenamefont {Guitteny}\ \emph {et~al.}(2015)\citenamefont
  {Guitteny}, \citenamefont {Mirebeau}, \citenamefont {Dalmas~de R\'eotier},
  \citenamefont {Colin}, \citenamefont {Bonville}, \citenamefont {Porcher},
  \citenamefont {Grenier}, \citenamefont {Decorse},\ and\ \citenamefont
  {Petit}}]{Guitteny2015}%
  \BibitemOpen
  \bibfield  {author} {\bibinfo {author} {\bibfnamefont {S.}~\bibnamefont
  {Guitteny}}, \bibinfo {author} {\bibfnamefont {I.}~\bibnamefont {Mirebeau}},
  \bibinfo {author} {\bibfnamefont {P.}~\bibnamefont {Dalmas~de R\'eotier}},
  \bibinfo {author} {\bibfnamefont {C.~V.}\ \bibnamefont {Colin}}, \bibinfo
  {author} {\bibfnamefont {P.}~\bibnamefont {Bonville}}, \bibinfo {author}
  {\bibfnamefont {F.}~\bibnamefont {Porcher}}, \bibinfo {author} {\bibfnamefont
  {B.}~\bibnamefont {Grenier}}, \bibinfo {author} {\bibfnamefont
  {C.}~\bibnamefont {Decorse}}, \ and\ \bibinfo {author} {\bibfnamefont
  {S.}~\bibnamefont {Petit}},\ }\href {\doibase 10.1103/PhysRevB.92.144412}
  {\bibfield  {journal} {\bibinfo  {journal} {Phys. Rev. B}\ }\textbf {\bibinfo
  {volume} {92}},\ \bibinfo {pages} {144412} (\bibinfo {year}
  {2015})}\BibitemShut {NoStop}%
\bibitem [{\citenamefont {Jensen}\ and\ \citenamefont
  {Mackintosh}(1991)}]{Jensen91}%
  \BibitemOpen
  \bibfield  {author} {\bibinfo {author} {\bibfnamefont {J.}~\bibnamefont
  {Jensen}}\ and\ \bibinfo {author} {\bibfnamefont {A.~R.}\ \bibnamefont
  {Mackintosh}},\ }\href@noop {} {\emph {\bibinfo {title} {Rare Earth
  Magnetism}}}\ (\bibinfo  {publisher} {Clarendon Press, Oxford},\ \bibinfo
  {year} {1991})\BibitemShut {NoStop}%
\bibitem [{\citenamefont {{Kadowaki}}\ \emph {et~al.}(2015)\citenamefont
  {{Kadowaki}}, \citenamefont {{Takatsu}}, \citenamefont {{Taniguchi}},
  \citenamefont {{F{\aa}k}},\ and\ \citenamefont {{Ollivier}}}]{Kadowaki2015}%
  \BibitemOpen
  \bibfield  {author} {\bibinfo {author} {\bibfnamefont {H.}~\bibnamefont
  {{Kadowaki}}}, \bibinfo {author} {\bibfnamefont {H.}~\bibnamefont
  {{Takatsu}}}, \bibinfo {author} {\bibfnamefont {T.}~\bibnamefont
  {{Taniguchi}}}, \bibinfo {author} {\bibfnamefont {B.}~\bibnamefont
  {{F{\aa}k}}}, \ and\ \bibinfo {author} {\bibfnamefont {J.}~\bibnamefont
  {{Ollivier}}},\ }\href {\doibase 10.1142/S2010324715400032} {\bibfield
  {journal} {\bibinfo  {journal} {SPIN}\ }\textbf {\bibinfo {volume} {05}},\
  \bibinfo {eid} {1540003} (\bibinfo {year} {2015})}\BibitemShut {NoStop}%
\bibitem [{\citenamefont {{Takatsu}}\ \emph {et~al.}(2017)\citenamefont
  {{Takatsu}}, \citenamefont {{Taniguchi}}, \citenamefont {{Kittaka}},
  \citenamefont {{Sakakibara}},\ and\ \citenamefont
  {{Kadowaki}}}]{Takatsu2017}%
  \BibitemOpen
  \bibfield  {author} {\bibinfo {author} {\bibfnamefont {H.}~\bibnamefont
  {{Takatsu}}}, \bibinfo {author} {\bibfnamefont {T.}~\bibnamefont
  {{Taniguchi}}}, \bibinfo {author} {\bibfnamefont {S.}~\bibnamefont
  {{Kittaka}}}, \bibinfo {author} {\bibfnamefont {T.}~\bibnamefont
  {{Sakakibara}}}, \ and\ \bibinfo {author} {\bibfnamefont {H.}~\bibnamefont
  {{Kadowaki}}},\ }\href {\doibase 10.1088/1742-6596/828/1/012007} {\bibfield
  {journal} {\bibinfo  {journal} {J. Phys.: Conf. Series}\ }\textbf {\bibinfo
  {volume} {828}},\ \bibinfo {eid} {012007} (\bibinfo {year}
  {2017})}\BibitemShut {NoStop}%
\bibitem [{Fak({\natexlab{a}})}]{Fak2015}%
  \BibitemOpen
  \href@noop {} {}\bibinfo {note} {{}B. {F{\aa}k}, H.
  {Kadowaki}, J. {Ollivier} and M. {Wakita}. (2015). Quadrupole order of
  Tb$_{2+x}$Ti$_{2-x}$O$_{7+y}$. Institut Laue-Langevin (ILL)
  doi:10.5291/ILL-DATA.4-05-628.}\BibitemShut {Stop}%
\bibitem [{Fak({\natexlab{b}})}]{Fak2016}%
  \BibitemOpen
  \href@noop {} {}\bibinfo {note} {{}B. {F{\aa}k}, H.
  {Kadowaki}, and J. {Ollivier}. (2016). Quadrupole order of
  Tb$_{2+x}$Ti$_{2-x}$O$_{7+y}$. Institut Laue-Langevin (ILL)
  doi:10.5291/ILL-DATA.4-05-635.}\BibitemShut {Stop}%
\bibitem [{\citenamefont {Kadowaki}({\natexlab{a}})}]{Kadowaki2018github}%
  \BibitemOpen
  \bibfield  {author} {\bibinfo {author} {\bibfnamefont {H.}~\bibnamefont
  {Kadowaki}},\ }\href@noop {} {}\bibinfo {note}
  {{}https://github.com/kadowaki-h/AbsorptionFactorIN5;
  https://github.com/kadowaki-h/AbsorptionFactorAMATERAS.}\BibitemShut {Stop}%
\bibitem [{\citenamefont {Ewings}\ \emph {et~al.}(2016)\citenamefont {Ewings},
  \citenamefont {Buts}, \citenamefont {Le}, \citenamefont {van Duijn},
  \citenamefont {Bustinduy},\ and\ \citenamefont {Perring}}]{Horace2016}%
  \BibitemOpen
  \bibfield  {author} {\bibinfo {author} {\bibfnamefont {R.}~\bibnamefont
  {Ewings}}, \bibinfo {author} {\bibfnamefont {A.}~\bibnamefont {Buts}},
  \bibinfo {author} {\bibfnamefont {M.}~\bibnamefont {Le}}, \bibinfo {author}
  {\bibfnamefont {J.}~\bibnamefont {van Duijn}}, \bibinfo {author}
  {\bibfnamefont {I.}~\bibnamefont {Bustinduy}}, \ and\ \bibinfo {author}
  {\bibfnamefont {T.}~\bibnamefont {Perring}},\ }\href {\doibase
  http://dx.doi.org/10.1016/j.nima.2016.07.036} {\bibfield  {journal} {\bibinfo
   {journal} {Nucl. Instrum. Methods Phys. Res. Sect. A}\ }\textbf {\bibinfo
  {volume} {834}},\ \bibinfo {pages} {132 } (\bibinfo {year}
  {2016})}\BibitemShut {NoStop}%
\bibitem [{\citenamefont {den Hertog}\ and\ \citenamefont
  {Gingras}(2000)}]{Hertog00}%
  \BibitemOpen
  \bibfield  {author} {\bibinfo {author} {\bibfnamefont {B.~C.}\ \bibnamefont
  {den Hertog}}\ and\ \bibinfo {author} {\bibfnamefont {M.~J.~P.}\ \bibnamefont
  {Gingras}},\ }\href {\doibase 10.1103/PhysRevLett.84.3430} {\bibfield
  {journal} {\bibinfo  {journal} {Phys. Rev. Lett.}\ }\textbf {\bibinfo
  {volume} {84}},\ \bibinfo {pages} {3430} (\bibinfo {year}
  {2000})}\BibitemShut {NoStop}%
\bibitem [{Sup()}]{Supplemental_Material}%
  \BibitemOpen
  \href@noop {} {}\bibinfo {note} {See Supplemental Material at https for
  further details of the least squares fits.}\BibitemShut {Stop}%
\bibitem [{\citenamefont {Li}\ \emph {et~al.}(2013)\citenamefont {Li},
  \citenamefont {Zhao}, \citenamefont {Fan}, \citenamefont {Zhang},
  \citenamefont {Zhou}, \citenamefont {Zhao},\ and\ \citenamefont
  {Sun}}]{Li2013}%
  \BibitemOpen
  \bibfield  {author} {\bibinfo {author} {\bibfnamefont {Q.~J.}\ \bibnamefont
  {Li}}, \bibinfo {author} {\bibfnamefont {Z.~Y.}\ \bibnamefont {Zhao}},
  \bibinfo {author} {\bibfnamefont {C.}~\bibnamefont {Fan}}, \bibinfo {author}
  {\bibfnamefont {F.~B.}\ \bibnamefont {Zhang}}, \bibinfo {author}
  {\bibfnamefont {H.~D.}\ \bibnamefont {Zhou}}, \bibinfo {author}
  {\bibfnamefont {X.}~\bibnamefont {Zhao}}, \ and\ \bibinfo {author}
  {\bibfnamefont {X.~F.}\ \bibnamefont {Sun}},\ }\href {\doibase
  10.1103/PhysRevB.87.214408} {\bibfield  {journal} {\bibinfo  {journal} {Phys.
  Rev. B}\ }\textbf {\bibinfo {volume} {87}},\ \bibinfo {pages} {214408}
  (\bibinfo {year} {2013})}\BibitemShut {NoStop}%
\bibitem [{\citenamefont {Kermarrec}\ \emph {et~al.}(2015)\citenamefont
  {Kermarrec}, \citenamefont {Maharaj}, \citenamefont {Gaudet}, \citenamefont
  {Fritsch}, \citenamefont {Pomaranski}, \citenamefont {Kycia}, \citenamefont
  {Qiu}, \citenamefont {Copley}, \citenamefont {Couchman}, \citenamefont
  {Morningstar}, \citenamefont {Dabkowska},\ and\ \citenamefont
  {Gaulin}}]{Kermarrec2015}%
  \BibitemOpen
  \bibfield  {author} {\bibinfo {author} {\bibfnamefont {E.}~\bibnamefont
  {Kermarrec}}, \bibinfo {author} {\bibfnamefont {D.~D.}\ \bibnamefont
  {Maharaj}}, \bibinfo {author} {\bibfnamefont {J.}~\bibnamefont {Gaudet}},
  \bibinfo {author} {\bibfnamefont {K.}~\bibnamefont {Fritsch}}, \bibinfo
  {author} {\bibfnamefont {D.}~\bibnamefont {Pomaranski}}, \bibinfo {author}
  {\bibfnamefont {J.~B.}\ \bibnamefont {Kycia}}, \bibinfo {author}
  {\bibfnamefont {Y.}~\bibnamefont {Qiu}}, \bibinfo {author} {\bibfnamefont
  {J.~R.~D.}\ \bibnamefont {Copley}}, \bibinfo {author} {\bibfnamefont
  {M.~M.~P.}\ \bibnamefont {Couchman}}, \bibinfo {author} {\bibfnamefont
  {A.~O.~R.}\ \bibnamefont {Morningstar}}, \bibinfo {author} {\bibfnamefont
  {H.~A.}\ \bibnamefont {Dabkowska}}, \ and\ \bibinfo {author} {\bibfnamefont
  {B.~D.}\ \bibnamefont {Gaulin}},\ }\href {\doibase
  10.1103/PhysRevB.92.245114} {\bibfield  {journal} {\bibinfo  {journal} {Phys.
  Rev. B}\ }\textbf {\bibinfo {volume} {92}},\ \bibinfo {pages} {245114}
  (\bibinfo {year} {2015})}\BibitemShut {NoStop}%
\bibitem [{\citenamefont {{Fennell}}\ \emph {et~al.}(2009)\citenamefont
  {{Fennell}}, \citenamefont {{Deen}}, \citenamefont {{Wildes}}, \citenamefont
  {{Schmalzl}}, \citenamefont {{Prabhakaran}}, \citenamefont {{Boothroyd}},
  \citenamefont {{Aldus}}, \citenamefont {{McMorrow}},\ and\ \citenamefont
  {{Bramwell}}}]{Fennell2009}%
  \BibitemOpen
  \bibfield  {author} {\bibinfo {author} {\bibfnamefont {T.}~\bibnamefont
  {{Fennell}}}, \bibinfo {author} {\bibfnamefont {P.~P.}\ \bibnamefont
  {{Deen}}}, \bibinfo {author} {\bibfnamefont {A.~R.}\ \bibnamefont
  {{Wildes}}}, \bibinfo {author} {\bibfnamefont {K.}~\bibnamefont
  {{Schmalzl}}}, \bibinfo {author} {\bibfnamefont {D.}~\bibnamefont
  {{Prabhakaran}}}, \bibinfo {author} {\bibfnamefont {A.~T.}\ \bibnamefont
  {{Boothroyd}}}, \bibinfo {author} {\bibfnamefont {R.~J.}\ \bibnamefont
  {{Aldus}}}, \bibinfo {author} {\bibfnamefont {D.~F.}\ \bibnamefont
  {{McMorrow}}}, \ and\ \bibinfo {author} {\bibfnamefont {S.~T.}\ \bibnamefont
  {{Bramwell}}},\ }\href {\doibase 10.1126/science.1177582} {\bibfield
  {journal} {\bibinfo  {journal} {Science}\ }\textbf {\bibinfo {volume}
  {326}},\ \bibinfo {pages} {415} (\bibinfo {year} {2009})}\BibitemShut
  {NoStop}%
\bibitem [{Mol()}]{Molavian2009}%
  \BibitemOpen
  \href@noop {} {}\bibinfo {note} {{H.~R. Molavian}, {P.~A. McClarty}, and
  {M.~J.~P. Gingras}, arXiv:0912.2957}\BibitemShut {NoStop}%
\bibitem [{Rau()}]{Rau_Gingras2018}%
  \BibitemOpen
  \href@noop {} {}\bibinfo {note} {{J.~G. Rau} and {M.~J.~P. Gingras},
  arXiv:1806.09638}\BibitemShut {NoStop}%
\bibitem [{NIS()}]{NISTdisclaimer}%
  \BibitemOpen
  \href@noop {} {}\bibinfo {note} {The identification of any commercial product
  or trade name does not imply endorsement or recommendation by the National
  Institute of Standards and Technology.}\BibitemShut {Stop}%
\bibitem [{\citenamefont {Kao}\ \emph {et~al.}(2003)\citenamefont {Kao},
  \citenamefont {Enjalran}, \citenamefont {Del~Maestro}, \citenamefont
  {Molavian},\ and\ \citenamefont {Gingras}}]{Kao03}%
  \BibitemOpen
  \bibfield  {author} {\bibinfo {author} {\bibfnamefont {Y.-J.}\ \bibnamefont
  {Kao}}, \bibinfo {author} {\bibfnamefont {M.}~\bibnamefont {Enjalran}},
  \bibinfo {author} {\bibfnamefont {A.}~\bibnamefont {Del~Maestro}}, \bibinfo
  {author} {\bibfnamefont {H.~R.}\ \bibnamefont {Molavian}}, \ and\ \bibinfo
  {author} {\bibfnamefont {M.~J.~P.}\ \bibnamefont {Gingras}},\ }\href
  {\doibase 10.1103/PhysRevB.68.172407} {\bibfield  {journal} {\bibinfo
  {journal} {Phys. Rev. B}\ }\textbf {\bibinfo {volume} {68}},\ \bibinfo
  {pages} {172407} (\bibinfo {year} {2003})}\BibitemShut {NoStop}%
\bibitem [{\citenamefont {Kadowaki}({\natexlab{b}})}]{Kadowaki2018githubLS}%
  \BibitemOpen
  \bibfield  {author} {\bibinfo {author} {\bibfnamefont {H.}~\bibnamefont
  {Kadowaki}},\ }\href@noop {} {}\bibinfo {note}
  {{}https://github.com/kadowaki-h/least65OMP.}\BibitemShut {Stop}%
\end{thebibliography}%
\end{document}